\documentclass[aps,pra,reprint,floatfix,showkeys]{revtex4-1}
\usepackage{amsmath,dcolumn,bm,siunitx}
\usepackage{graphicx,color}
\usepackage{physics}
\usepackage{placeins,etoolbox}
\makeatletter
\patchcmd\linenumberpar{\@LN@parpgbrk}{\penalty\@LN@parpgpen\relax}{}{}
\makeatother

\usepackage[caption=false]{subfig}
\DeclareGraphicsExtensions{{.pdf},{.jpg}}
\graphicspath{{figures/jpg/}}

\newcommand*\chem[1]{\ensuremath{\mathrm{#1}}}
\newcommand*\chembf[1]{\ensuremath{\mathbf{#1}}}
\renewcommand*\Tr{\ensuremath{\mathrm{Tr}}}
\renewcommand*\vec[1]{\textbf{#1}}
\begin{document}

\title{Monte Carlo Simulations of Spin Transport in a Strained Nanoscale InGaAs Field Effect Transistor}

\author{B.~Thorpe}
\email[]{ben4195@gmail.com}
\affiliation{Department of Physics, College of Science, Swansea University, Singleton Park, Swansea, SA2 8PP, United Kingdom}

\author{K.~Kalna}
\homepage{http://engweb.swan.ac.uk/~karolkalna/}
\email[]{k.kalna@swansea.ac.uk}
\affiliation{Electronic Systems Design Centre, College of Engineering, Swansea University, Bay Campus, Fabian Way, Swansea, SA1 8EN, United Kingdom}

\author{F.~C.~Langbein}
\email{frank@langbein.org}
\homepage{https://langbein.org/}
\affiliation{School of Computer Science \& Informatics, Cardiff University, 5 The Parade, Cardiff, CF24 3AA, United Kingdom}

\author{S.~Schirmer}
\email{sgs29@swan.ac.uk,lw1660@gmail.com}
\homepage{http://quantumcontrol.info/}
\affiliation{Department\ of Physics, College of Science, Swansea University, Singleton Park, Swansea, SA2 8PP, United Kingdom}

\date{\today}

\begin{abstract}
 Spin-based logic devices could operate at very high speed with very low energy consumption and hold significant promise for quantum information processing and metrology. Here, an in-house developed, experimentally verified, ensemble self-consistent Monte Carlo device simulator with a Bloch equation model using a spin-orbit interaction Hamiltonian accounting for Dresselhaus and Rashba couplings is developed and applied to a spin field effect transistor (spinFET) operating under externally applied voltages on a gate and a drain.  In particular, we simulate electron spin transport in a \SI{25}{nm} gate length \chem{In_{0.7}Ga_{0.3}As} metal-oxide-semiconductor field-effect transistor (MOSFET) with a CMOS compatible architecture.  We observe non-uniform decay of the net magnetization between the source and gate and a magnetization recovery effect due to spin refocusing induced by a high electric field between the gate and drain. We demonstrate coherent control of the polarization vector of the drain current via the source-drain and gate voltages, and show that the magnetization of the drain current is strain-sensitive and can be increased twofold by strain induced into the channel.
\end{abstract}


\maketitle

\section{\label{sec:intro}Introduction}

Spin is one of the most intriguing quantum properties carried by elementary particles. Incorporation of electron spin into the operation of semiconductor devices enables novel functionality and increased performance for information processing and metrology~\cite{Awschalom2013,Wolf2001}.  Among the most promising spin-based semiconductor devices is the spin field effect transistor (spinFET) \cite{Datta1990}, considered a candidate for future high performance digital computing and memory with ultra low energy needs~\cite{ITRS2013}.

The original spinFET proposal~\cite{Datta1990} relied on a simple ballistic transport model to argue that the source drain current in a FET-like structure with ferromagnetic source and drain electrodes will be modulated by the  voltage applied to the gate electrode as a result of Rashba spin-orbit coupling.  Gate control of spin transport along the channel has been soon recognised as the major challenge. A striped-channel high electron mobility transistor (HEMT)~\cite{Bournel1997} with a quasi-1D channel was proposed using ensemble Monte Carlo simulations of electron transport in quantum-mechanically confined channel to minimise spin dephasing by confining Rashba spin precession to fewer dimensions~\cite{Bournel1998, Bournel2000, Bournel2001}. The ensemble Monte Carlo simulations were to used to predict spin coherence and transconductance effects in the 2D quantum-confined channel of a heterostructure III-V HEMT assuming one subband approximation~\cite{MinShen2004}. However, the simulations did not use a realistic gate model and were not self-consistently coupled with the Poisson equation thus the results are valid only at low applied electric field~\cite{MinShen2004}. Recognizing spin dephasing in a 2D channel as a limiting factor for a spinFET, \cite{Schliemann2003} proposed to eliminate dephasing by tuning Rashba and Dresselhaus couplings such that the effects of spin orbit coupling effectively cancel and the model becomes analytically solvable.  Although tuning of Rashba and Dresselhaus coupling has been demonstrated experimentally~\cite{PRL107n136604}, these experiments involved complex multiple quantum well structures not compatible with typical 2D channel FETs, for which the Rashba effect is generally recognized as dominant over Dresselhaus coupling in typical III-V FET~\cite{Bournel1998}.

\begin{figure*}[t]
  \centering
  \includegraphics[width=.8\textwidth]{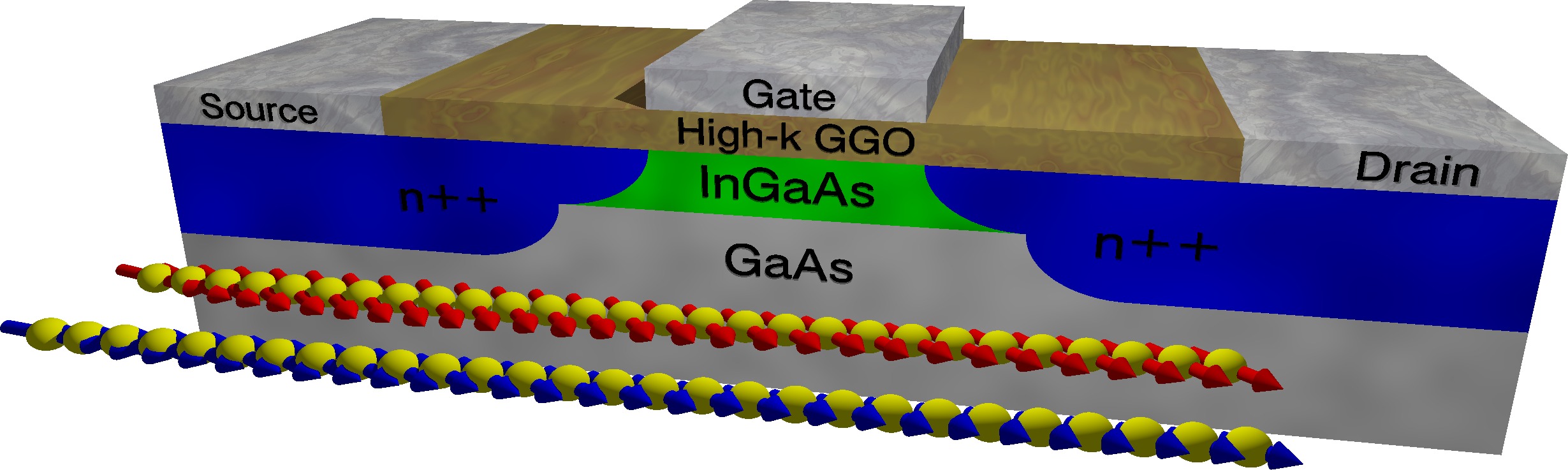}
  \caption{\label{3D}3D model of the studied \chem{In_{0.3}Ga_{0.7}As} MOSFET showing spin polarization of electrons along $n$-channel with $4\%$ strain in the $[001]$ direction (Red) and unstrained (Dark Blue).}
\end{figure*}

The objective of this paper is to study non-equilibrium spin transport in a realistic \SI{25}{nm} gate length \chem{In_{0.3}Ga_{0.7}As} FET as shown in Fig.~\ref{3D} at room temperature by incorporating the electron spin degrees of freedom into self-consistent semi-classical transport device simulations. These device simulations use finite-element quantum-corrected ensemble Monte Carlo technique self-consistently coupled to solutions of Poisson equation~\cite{Kalna2008,N.Seoane2016}. The coupling physically means that long-range electron-electron interactions are included in the simulations \cite{Kalna2008}. The device geometry studied is identical to the architecture of planar MOSFETs based on an \chem{In_{0.3}Ga_{0.7}As} channel on a Si substrate~\cite{Takagi2014}, the only modification being that the source and drain gates are assumed to be ferromagnetic.  Although the aforementioned simulation results are clearly not applicable to our device, one might conjecture that the spin modulation effect for nanoscale devices with channel lengths on the order of \SI{50}{nm} at room temperature would be negligible considering spatial modulation lengths on the order of microns predicted in earlier work.  However, our simulation results suggest that, although reduced, the spin precession effect should still be observable.  In particular, the polarization of the electrons initially decays as expected as they traverse the device, but partially recovers as the electrons approach the drain.  It appears that the electron spins are initially dephased but then partially refocused by the electric field at the gate electrode.  The recovery of the magnetization is present independent of the polarization of the drain electrode and can therefore not be attributed to existing polarized carriers inside the drain.  Moreover, the decay and recovery depend on the gate voltage and remain controllable.  Our simulator also enables us to study the effect of strain on the spin transport. Since Dresselhaus and Rashba effects depend on the direction and strength of the strain in the device, electron polarization at the drain of the device changes as a result of strain.

This paper is organized as follows. In Section~\ref{sec:theory}, we briefly describe the simulations and the underlying theory. In Section~\ref{sec:nostrain}, we present and discuss the results of spin transport simulations for an unstrained \chem{In_{0.3}Ga_{0.7}As} MOSFET device including steady-state magnetization across the channel and the effect of the source-drain and gate voltages, respectively.  In Section~\ref{sec:strain}, the effect of microscopic strain is investigated based on $\vec{k}\cdot\vec{p}$ calculations to determine the strain dependence of the spin-orbit couplings and its effect on spin transport and the observed magnetization of the drain current.

\section{\label{sec:theory}Electron Spin Monte Carlo}

The main simulation technique used in this work to model non-equilibrium many-body electron transport in realistic semiconductor devices is the ensemble Monte Carlo (MC) technique~\cite{Tomizawa1993}, self-consistently coupled with solutions of the Poisson equation, accounting for both long-range electron-electron interactions and quantum corrections using the effective quantum potential~\cite{Ferry2000}, assembled in an in-house-developed 2D finite element heterostructure MC device simulation tool~\cite{AynulIslam2011,Kalna2008}. This device simulation tool was verified with experimental data for various devices, including measured I-V characteristics of a \SI{120}{nm} gate length \chem{In_{0.2}Ga_{0.8}As} pseudomorphic~\cite{Kalna2002}, lattice matched metamorphic HEMT~\cite{Moran2003}, and a \SI{50}{mn} gate length \chem{In_{0.7}Ga_{0.3}As}/\chem{InP} HEMT~\cite{Kalna2005}.

The MC engine uses an analytical band-structure with three anisotropic valleys ($\Gamma$, L and X) with non-parabolic energy dispersion~\cite{Kalna2002}.  Electron scattering with all essential scattering mechanisms present in III-V semiconductors is considered: polar optical phonons, inter-valley and intra-valley optical phonons, non-polar optical phonons, acoustic phonons, interface roughness, interface phonons at the dielectric/semiconductor interface, and ionized impurity scattering using static screening model. Furthermore, the alloy scattering as well as strain effects on bandgap, electron effective mass, optical phonon deformation potential and energy are included in the device channel~\cite{AynulIslam2011}. Details of this particular MC device simulation tool can be found in~\cite{Benbakhi2012,Kalna2008,Kalna2005} while details on the ensemble MC technique are in textbooks~\cite{C.Jacoboni1989,Tomizaw1993}. The electron spin is treated in the ensemble MC technique as a separate degree of freedom of the electrons using the spin density matrix $\rho_0(t)$~\cite{MinShen2004} in addition to already present electron description in 3D $\mathbf{k}$-space and 2D real-space.

The spin state of a spin-$\tfrac{1}{2}$ particle such as an electron can be described by the density matrix
\begin{equation}\label{denmat}
  \rho_0(t)= \begin{pmatrix}
                \rho_{\uparrow \uparrow}(t)   & \rho_{\uparrow \downarrow}(t) \\
                \rho_{\downarrow \uparrow}(t) & \rho_{\downarrow \downarrow}(t)
             \end{pmatrix},
\end{equation}
where $\rho_{\downarrow \uparrow}(t)=\rho_{\uparrow \downarrow}(t)^*$ and $\rho_{\uparrow \uparrow}(t)+\rho_{\downarrow \downarrow}(t)=1$.  $\rho_{\uparrow \uparrow}$ and $\rho_{\downarrow \downarrow}$ represent the probability of finding the electron in either a spin up or spin down state  and $\rho_{\uparrow \downarrow}$, $\rho_{\downarrow \uparrow}$ represent the coherence.

\begin{figure}
  \centering
  \includegraphics[width=.6\columnwidth]{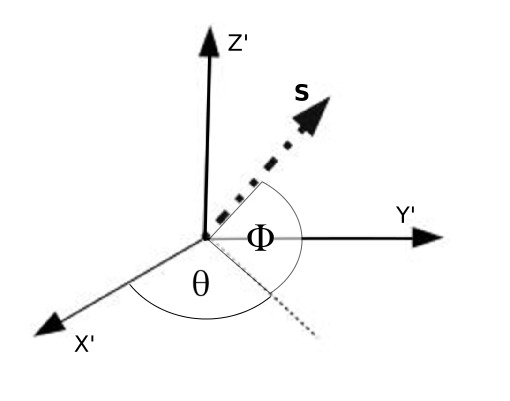}
  \caption{\label{rot_axis}Diagram showing the azimuthal ($\theta$) and elevation ($\phi$) angles used for the rotational analysis. The injection direction is mapped to the $x'$ axis.}
\end{figure}

The evolution of the spin states of individual electrons is governed by $\rho(t) = U(t) \rho_0 U(t)^\dag$ where $U(t)$ is unitary propagator satisfying the Schr\"odinger equation
\begin{equation} \label{eq:SE}
  i\hbar\frac{d}{dt} U(t) = H U(t), \qquad U(0)=I.
\end{equation}
Here $\hbar$ is the reduced Planck constant and $H$ is the Hamiltonian operator of the system, which we take to be a spin-orbit interaction Hamiltonian consisting mainly of two terms: (i) the simplified Dresselhaus Hamiltonian~\cite{MinShen2004}
\begin{equation}\label{dress}
  H_D = \gamma \langle k_y^2 \rangle (k_z\sigma_z-k_x\sigma_x),  \quad k_x^2,k_z^2 \ll \langle k_y^2\rangle,
\end{equation}
which accounts for spin-orbit coupling as a result of bulk inversion asymmetry of the crystal, and (ii) the Rashba Hamiltonian
\begin{equation}\label{Rash}
  H_R = \alpha_{br}(k_z\sigma_x-k_x\sigma_z),
\end {equation}
which accounts for spin-orbit coupling due to structural inversion asymmetry of the quantum well. Here $x$ is taken to be the transport direction along the device channel and $y$ the growth direction of the quantum well.
Discretizing the equations, we obtain the update rule for the density matrix,
\begin{equation}\label{matstep}
  \rho(t+\tau)=e^{-i(H_R+H_D)\tau/\hbar} \rho(t)e^{i(H_R+H_D)\tau/\hbar}.
\end{equation}
Using basic matrix algebra it can easily be shown that
\begin{equation}
  e^{-i(H_R+H_D)\tau/\hbar} =
  \begin{pmatrix}
    \cos{(|\alpha|\tau)}                                           & i\frac{\alpha}{|\alpha|} \sin{(|\alpha|\tau)} \\
    i\frac{\alpha^*}{|\alpha|} \sin{(|\alpha|\tau)} & \cos{(|\alpha|\tau)}
  \end{pmatrix}
\end{equation}
with
\begin{equation}
  \alpha = \hbar^{-1} [(\alpha_{br} k_z - \gamma \langle k_y^2\rangle k_x)+i(\alpha_{br} k_x - \gamma \langle k_y^2\rangle k_z)].
\end{equation}
This shows that the evolution of the spin polarization vector is equivalent to a rotation determined by the direction of the electron momentum. 
The spin-orbit interaction for a single electron spin is Hamiltonian (see Eqs.~\eqref{Rash} and \eqref{dress}) and hence effects a coherent rotation.  However,  as the axis and angle of this rotation depend on the $\vec{k}$-vector, i.e., momentum, of the electron and the electrons have different $\vec{k}$-vectors due to scattering and ballistic transport, they experience different rotations.  Thus, even if all electron spins initially point in the same direction, the different rotations they undergo quickly cause the electron spins to point in different directions, resulting in a decrease or loss of net magnetization of the ensemble of electrons in the channel.  However, if a large number of electrons undergo similar rotations, as may be the case in a strong electric field, the result will be a net rotation of the magnetization vector of the electron ensemble.  To facilitate comparison of coherent rotations of the magnetization vector during transport independent of the injection polarization, we map the Bloch vector for each injection case onto a common set of axes (denoted as $x'$, $y'$ and $z'$) and use spherical polar coordinates to describe the relative orientations as shown in Fig.~\ref{rot_axis}. The injection direction is mapped to the $x'$ axis in all cases while orthogonal directions define the azimuthal ($\theta$) and elevation ($\phi$) angles.  

Our model neglects spin flips due to impurity or phonon scattering, the Elliott-Yafet mechanism, and spin relaxation caused by hyperfine coupling, based on the observation that Dyakonov-Perel mechanism is the dominant source of spin relaxation in \chem{GaAs}~\cite{JaroslavFabian2007} but such effects could easily be incorporated into the simulator.

The spin polarization vector $\vec{s}=(s_x,s_y,s_z)^T$ is obtained from $s_\zeta(t)=\Tr(\sigma_\zeta \rho(t))$, where $\sigma_\zeta$ for $\zeta =x,y,z$ are the Pauli matrices
\begin{equation}
  \label{eq:pauli}
  \sigma_x = \begin{pmatrix} 0 & 1\\  1 & 0 \end{pmatrix}, \quad
  \sigma_y = \begin{pmatrix} 0 & -i\\ i & 0 \end{pmatrix}, \quad
  \sigma_z = \begin{pmatrix} 1 & 0\\  0 & -1 \end{pmatrix}.
\end{equation}
The components of the spin polarization vector of the current $S_\zeta = \langle s_\zeta(r,t)\rangle$ are obtained by averaging the components $s_\zeta(r,t)$ of the spin polarization vectors of all electrons in a thin slice through the device channel (along the direction of transport) located at position $x=r$ at time $t$. The magnitude $\norm{\vec{s}(r,t)} \leq 1$ defines the amount of polarization, $1$ being defined as $100\%$ spin polarization in the direction of $\vec{s}(r,t)$.

\section{\label{sec:nostrain}Spin Transport in Nanoscale \chembf{In_{0.3}Ga_{0.7}As} MOSFET}

\begin{figure}
  \centering
  \includegraphics[width=\linewidth]{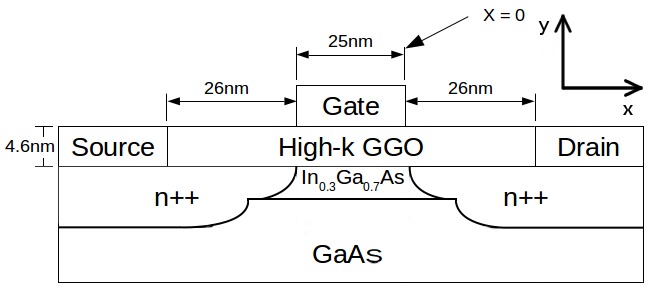}
  \caption{Cross section of the \SI{25}{nm} gate length, $n$-channel \chem{In_{0.3}Ga_{0.7}As} MOSFET.}
  \label{MOSFET}
\end{figure}

\begin{figure}
  \includegraphics[width=\columnwidth]{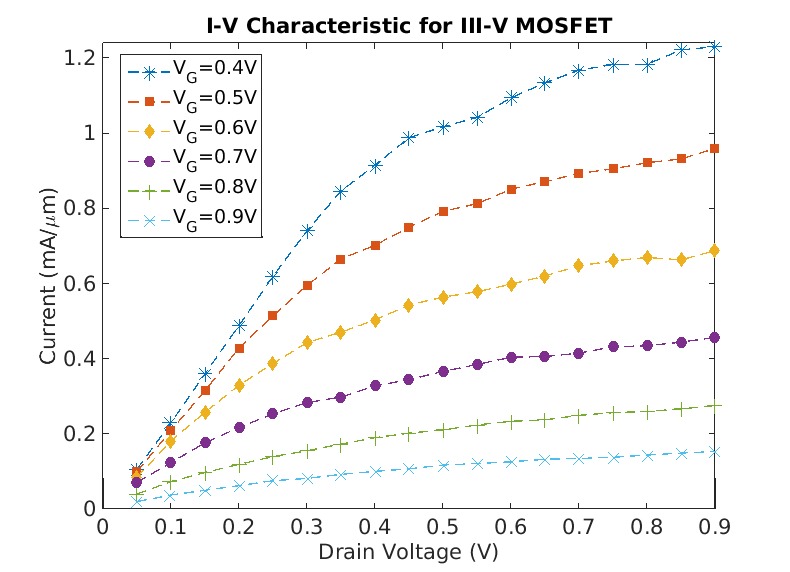}
  \caption{I$_\mathrm{D}$-V$_\mathrm{D}$ characteristic at indicated gate biases (V$_\mathrm{G}$) for the \SI{25}{mn} gate length \chem{In_{0.3}Ga_{0.7}As} MOSFET.}
  \label{fig:IV}
\end{figure}

\begin{figure*}
  \null\hfill
  \subfloat[$x$ component of magnetization]{\includegraphics[width=0.32\textwidth]{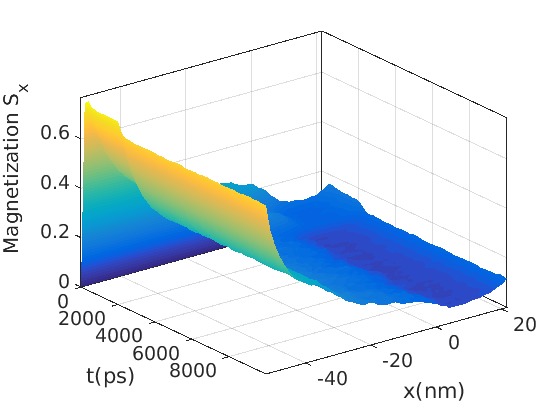}}\hfill
  \subfloat[$y$ component of magnetization]{\includegraphics[width=0.32\textwidth]{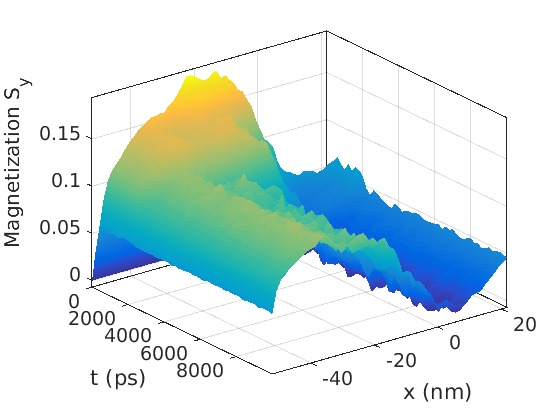}}\hfill
  \subfloat[$z$ component of magnetization]{\includegraphics[width=0.32\textwidth]{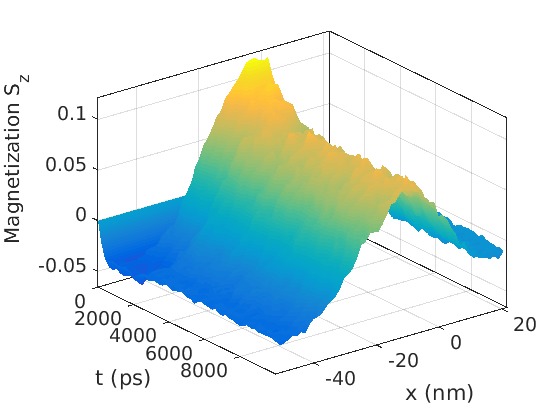}}\hfill\null\\[-2ex]
  \caption{\label{mag_x_t} Spatio-temporal evolution of the components of the magnetization vector for injection of $S_x$-polarized spins with gate and source-drain voltages of \SI{0.9}{V} each.  $x$ is the position along the channel with $x=0$ being set at the end of the gate.}\vspace*{-3ex}
  \null\hfill
  \subfloat[Injection $S_x$-polarized.]{\label{mag_x_sx}\includegraphics[width=0.32\textwidth]{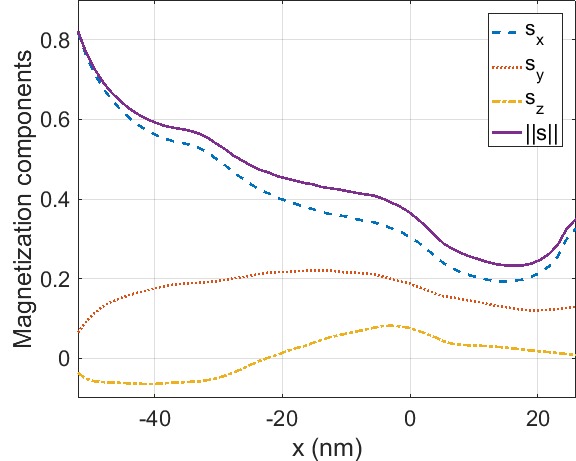}}\hfill
  \subfloat[Injection $S_y$-polarized.]{\label{mag_x_sy}\includegraphics[width=0.32\textwidth]{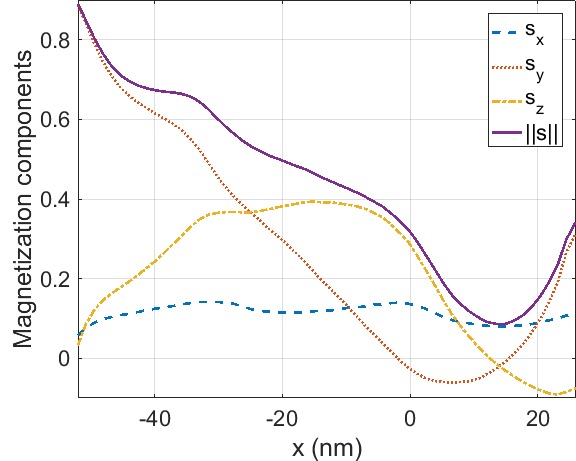}}\hfill
  \subfloat[Injection $S_z$-polarized.]{\label{mag_x_sz}\includegraphics[width=0.32\textwidth]{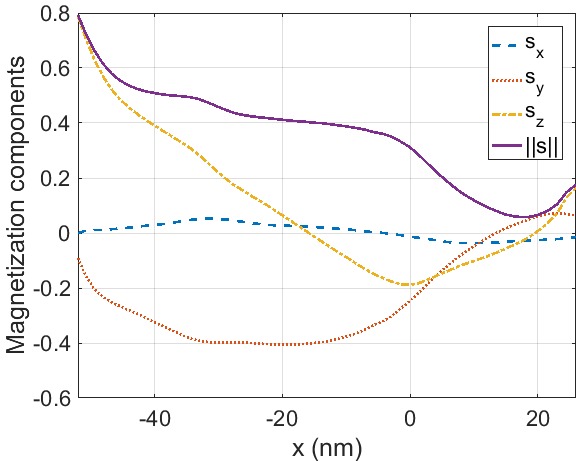}}\hfill\null\\[-2ex]
  \caption{\label{mag_x}Components of magnetization vector vs position along the channel (averaged over \num{10} runs) taken at $t=\SI{8}{ps}$, i.e., after a steady state was reached, for different injection polarizations at the same gate (V$_\mathrm{G}$) and source-drain voltage (V$_\mathrm{D}$) of \SI{0.9}{V}.}\vspace*{-3ex}
  \null\hfill
\end{figure*}

We simulate spin transport in the \SI{25}{nm} gate length \chem{In_{0.3}Ga_{0.7}As} MOSFET with a spacer of \SI{26}{nm} at room temperature ($T=\SI{300}{K}$). The cross-section of the transistor under study is shown in Fig.~\ref{MOSFET}. The device consists of a \SI{400}{nm} \chem{GaAs} substrate, a \SI{7}{nm} thick \chem{In_{0.3} Ga_{0.7}As} channel, a \SI{4.6}{nm} layer of high-$\kappa$  \chem{Ga_2 O_3 /(Gd _x Ga_{1-x} )_2 O_3} (GGO, $\kappa = 20$) separating the channel from a metal gate with a work function of \SI{4.05}{eV}. The structure has a background uniform $p$-type doping of \SI{1e18}{cm^{-3}} and $n$-type Gaussian-like doping in the S/D including extensions with a maximum doping of \SI{2e19}{cm^{-3}}. The I$_\mathrm{D}$-V$_\mathrm{D}$ (drain current vs. drain bias) characteristics of the \chem{In_{0.3}Ga_{0.7}As} MOSFET shown in Fig.~\ref{fig:IV} exhibit a large on-current of about $1.15$~mA/$\mu$m at overdrive of $0.7$~V (V$_\mathrm{D}$-V$_\mathrm{T}=0.7$~V, V$_\mathrm{D}=0.7$~V) having a threshold voltage (V$_\mathrm{T}$) of $0.2$~V as needed for digital applications~\cite{ITRS2013}.

The source of the device was assumed to be ferromagnetic such that electrons injected from the source into the channel would be spin polarized.  Several proposals for spin injection exist using ferromagnetic electrodes exists~\cite{PRL83p203}.  Electrical spin injection has recently also been demonstrated experimentally using quantum point contacts~\cite{PojenChuang2014} although the latter required sub-Kelvin temperatures, which is not compatible with room-temperature operation as assumed for our device.  The electrons inside the channel were initialised such that there was no net magnetization across the channel.  The simulation was then run with \num{100000} super-particles with gate voltages of \SIrange{0.5}{0.9}{V} and source-drain voltages of \SIrange{0.5}{0.9}{V}, respectively, in time steps of \SI{1}{fs} for a total time of \SI{10}{ps}. For each time step, the average polarization vector was calculated for the electrons contained in $100$ evenly spaced slices across the channel. The entire process was repeated for three different injected polarizations, realized by setting the spin-polarization vector to be parallel to the $x$, $y$ or $z$-axis upon injection into the channel from the source reservoir.

\begin{figure*}
  {\centering\includegraphics[width=.8\textwidth]{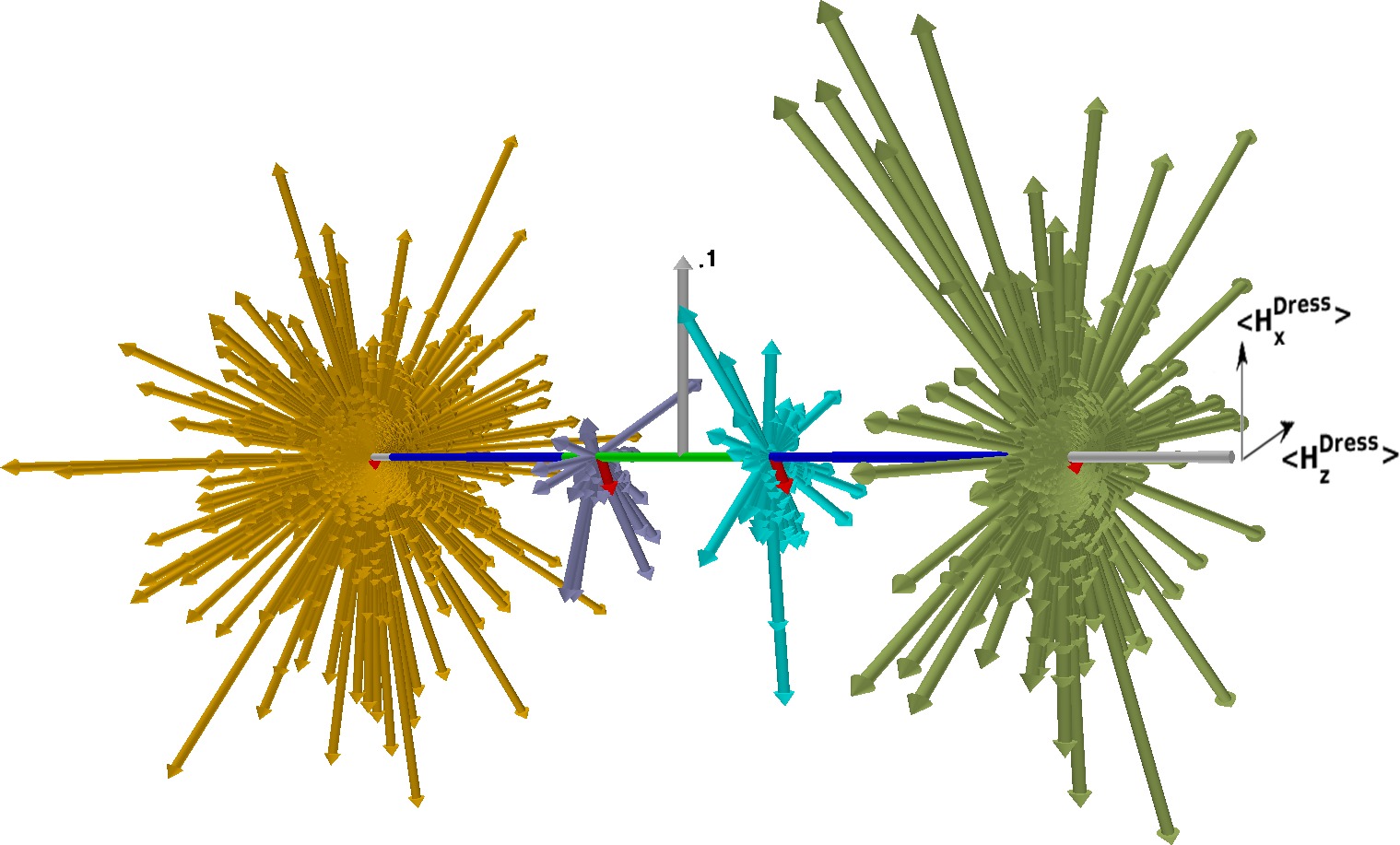}}
  \caption{Dresselhaus Hamiltonian vectors (in units of \si{meV}) of 4 electron ensembles corresponding to thin slices along the channel for a single Monte Carlo run ($V_\mathrm{G}=\SI{0.9}{V}, V_\mathrm{D}=\SI{0.5}{V}$) orange (far left) x=-55nm, RichBlue (centre-left) x=-20nm, Cyan (centre right) x=0nm, Forest Green (far right) x=27nm. Grey arrow show the scale whilst Red arrows show the average.}
  \label{fig:Dresselhaus-field}
\end{figure*}

\begin{figure}
  \includegraphics[width=\columnwidth]{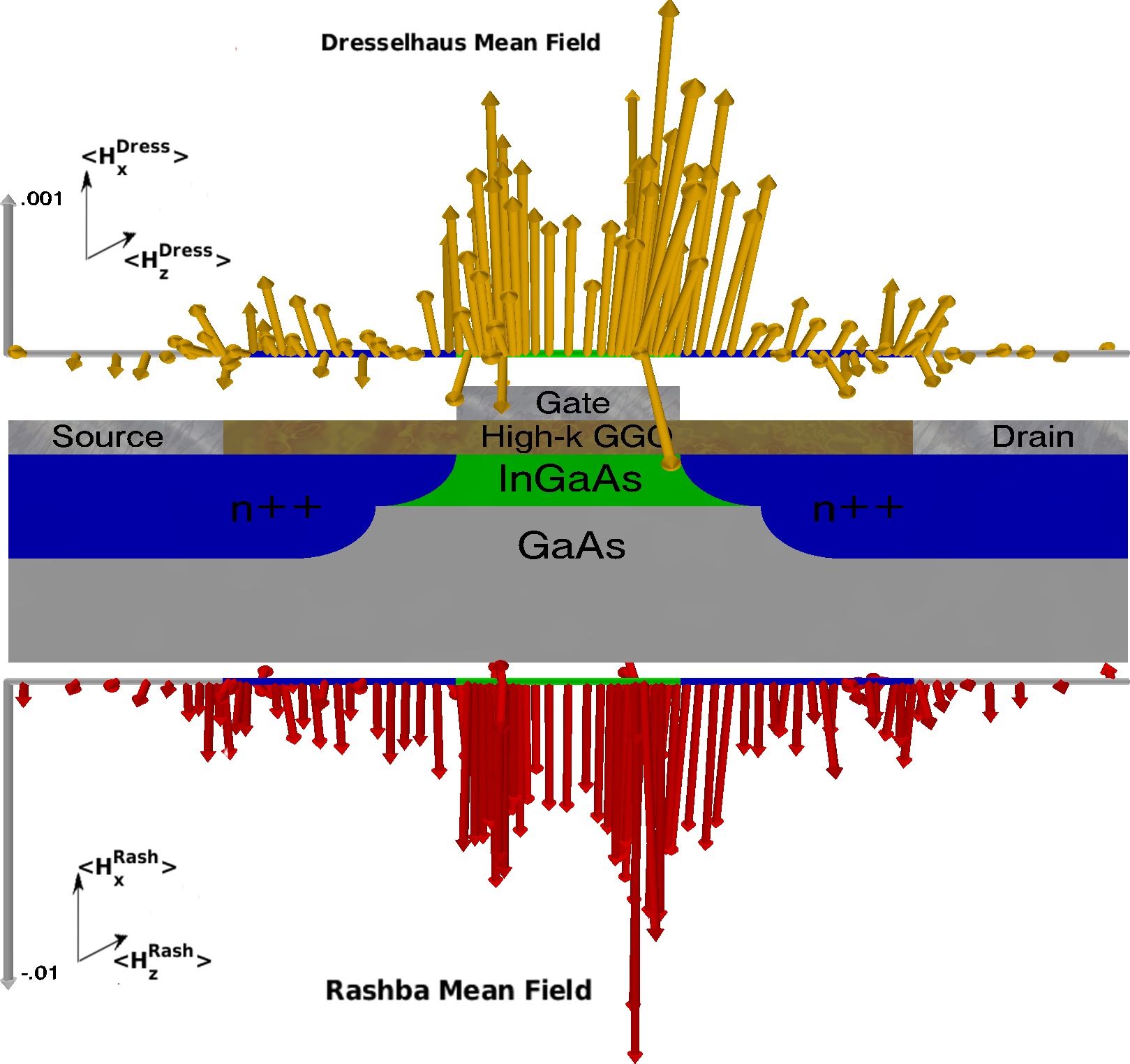}
  \caption{Rashba and Dresselhaus mean field vectors $\vec{H}_R$ and $\vec{H}_D$, obtained by averaging over all particles in thin slices across the channel for a single Monte Carlo run ($V_\mathrm{G}=\SI{0.9}{V}, V_\mathrm{D}=\SI{0.5}{V}$).  The $z$-axis is in plane perpendicular to the channel but for the vector plots the axes have been rotated so that $H_z$ is in the vertical direction for visual clarity.}
  \label{fig:mean-field-plot}
\end{figure}

\begin{figure}
  \includegraphics[width=\columnwidth]{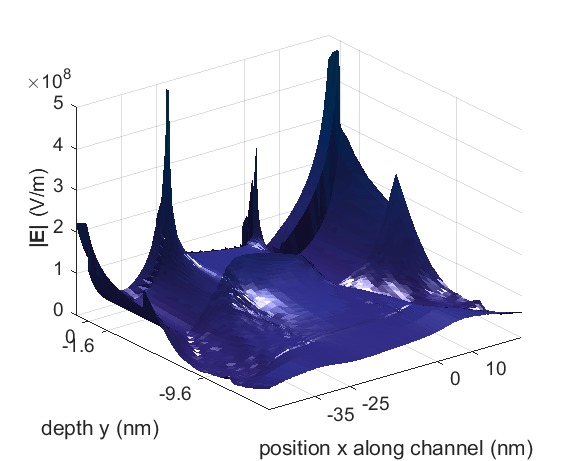}
  \caption{Steady-state electric field $\norm{\vec{E}}$ in the device for one run of Monte Carlo simulator with $V_G=\SI{0.9}{V}$ and $V_D=\SI{0.5}{V}$.  The electric field is non-uniform, concentrated in the channel and near the electrodes, with large fringe fields at electrode boundaries.}
  \label{fig:e-field-plot}
\end{figure}

As we are primarily interested in spin \emph{transport} in this paper, the process of spin injection was assumed to be $100\%$ efficient for the sake of simplicity.  This assumption is often used in spin transport simulations \cite{Bournel2000,Bournel1997,Bournel1998,A.Makarov2016,V.Sverdlov2015} and can be justified here as follows.  If the experimental realization of the spin injecting contact or interface provides a lower efficiency of spin injection, as expected in practice, then the simulation results can be adjusted by scaling them accordingly.  Specifically in our model, the spins of the injected electrons are all pointing in the same direction while all other electrons have spins pointing in random directions, resulting in no net magnetization.  Reducing the injection efficiency in this model is equivalent to reducing the initial length of the Bloch vector representing the net magnetization.  If the injection efficiency is $\eta<1$ then the initial state of system can be represented by the mixed-state density operator $\rho_0 = \eta |\uparrow\rangle\langle \uparrow| + (1-\eta)\rho_*$, where $\rho_*$ represents the completely mixed state with no net magnetization.  Scattering in the channel then further reduces the length of the Bloch vector but there is no reason to expect that spin relaxation or rotation effects will be affected by lower injection efficiency.  Therefore, lower injection efficiencies can be accounted for by simply mixing the Bloch vector of the polarized spins with the completely mixed state.

\subsection{Steady-State Magnetization along Channel}

Fig.~\ref{mag_x_t} shows that the components of the magnetization vector quickly (in about \SI{2000}{ps}) approach a steady state.  Fig.~\ref{mag_x} shows the variation of the components of the magnetization vector along the channel after a steady-state has been reached for three different injection polarizations.  In all three cases, the decay is not simply exponential, as might be expected based on simple drift-diffusion model simulations.  In particular, we observe a magnetization recovery as we approach the channel region at the drain side, which is most pronounced when the injection polarization is in the $S_y$-direction, i.e., in the growth direction.

For $S_x$-polarized injection (Fig.~\ref{mag_x_sx}) we see a high $S_x$-polarization at the source-channel boundary ($x=\SI{-52}{nm}$, which decreases non-uniformly as we cross the channel from the source to the drain before recovering slightly between the right edge of the gate ($x=\SI{0}{nm}$) and the drain, leading to a net magnetization at the left edge of the drain ($x=\SI{26}{nm}$) of $\vec{s}=(0.36,0.25,0.08)$.  Fig.~\ref{mag_x_sy} shows that the magnetization recovery is most prominent for the $S_y$ injection case where the magnetization rises from a minimum of $S_y=0.01$ at the right gate edge to $S_y=0.28$ at the left drain edge.

The component resolved plots also show that the polarization vector of the current undergoes a coherent rotation as we move along the channel. For instance, in Fig.~\ref{mag_x_sx}, the initial fall in the $S_x$-magnetization is accompanied by an increase in the $S_y$-component, and later an increase in the $S_z$-component. As no external magnetic fields are applied to directly rotate the spin polarization while the electrons are moving through the channel, this effect must be attributed to spin-orbit coupling.

To elucidate the spin-orbit coupling effect, we calculate the components of the Rasha and Dresselhaus spin-orbit coupling Hamiltonians \begin{align}
  \vec{H}_R^{(n)} &= \left( \Tr(H_R \sigma_x), \Tr(H_R \sigma_y), \Tr(H_R \sigma_z) \right) \nonumber \\
                             &= 2\alpha_{br} (k_z,0,-k_x), \nonumber \\
  \vec{H}_D^{(n)} &= \left( \Tr(H_D \sigma_x), \Tr(H_D \sigma_y), \Tr(H_D \sigma_z) \right) \nonumber \\
                             &= 2\gamma \langle k_y^2 \rangle (-k_x,0,k_z)
\end{align}
for all particles in the channel.  Fig.~\ref{fig:Dresselhaus-field} shows the distribution of the Dresselhaus field vectors $\vec{H}_D^{(n)}$ for all electrons (blue arrows) in a thin slice along the channel and the resulting mean field vector $\vec{H}_D = \langle \vec{H}_D^{(n)}\rangle$ averaged over all particles (bold red) in the slice for positions in the channel.  For $x\approx \SI{-55}{nm}$ the field vectors are randomly oriented, resulting in a vanishing mean field, while for $x\approx\SI{0}{nm}$ there is less spread in the individual field vectors and the mean field vector $\vec{H}_D$ does not vanish.  In the first case the electrons in the slice experience rotations about randomly oriented axes as a result of spin orbit coupling, while in the second case the rotations are not completely randomly oriented, resulting in a net coherent rotation of the electron ensemble.

The Rashba and Dresselhaus mean field vectors for electron ensembles in thin slices across the channel shown in Fig.~\ref{fig:mean-field-plot} suggest the Rashba and Dresselhaus mean fields are small, away from the gate, but quite large in the gate region.  This suggests that away from the gate spin-orbit coupling mostly imparts random kicks to the electron spins resulting in dephasing, while in the gate region the strong electric field of the gate causes the spin-orbit coupling to act more like a coherent rotation of the electron spins.  Although this magnetization recovery effect appears surprising, oscillatory behavior of the magnetization components has been observed in earlier work on spin transport in 2DEGs subject to a constant electric driving field across the channel.  For example, Bournel \emph{et al.} \cite{Bournel2001} observe oscillations in the spin polarization along the channel, albeit on different length scales for lower, uniform electric fields.  In both cases the electric field effects a coherent rotation of the net magnetization.  However, as the electric field generated by the gate electrode in the actual device simulated, shown in Fig.~\ref{fig:e-field-plot}, is concentrated in the gate region and highly non-uniform, we do not expect a simple oscillatory magnetization pattern.  Indeed, it may be possible to optimize the device design to maximize the recovery effect at the drain.

\subsection{Gate Voltage Dependence}

\begin{figure*}
  \null\hfill
  \subfloat[Total Magnetization]{\label{fig:total-vs-VG} \includegraphics[width=0.32\textwidth]{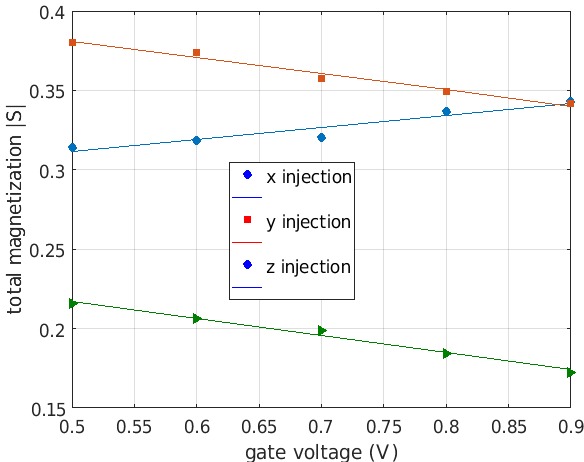}}\hfill
  \subfloat[$\theta$ angle]        {\label{fig:theta-vs-VG}\includegraphics[width=0.32\textwidth]{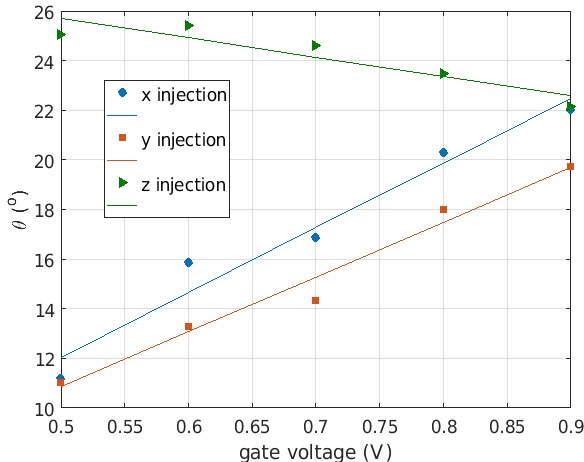}}\hfill
  \subfloat[$\phi$ angle]           {\label{fig:phi-vs-VG}    \includegraphics[width=0.32\textwidth]{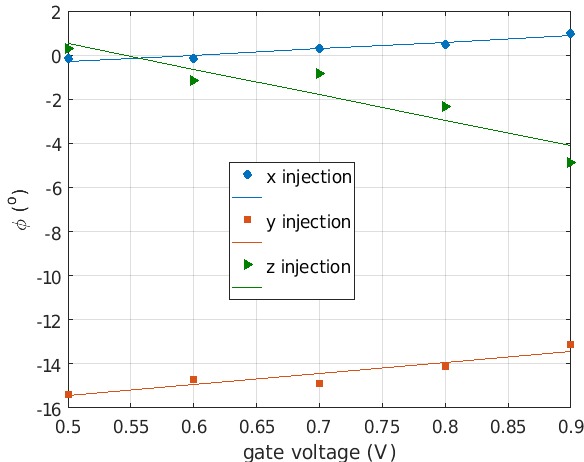}}\hfill\null\\
  \caption{\label{fig:VG-dependence}Magnetization at left drain edge ($x=\SI{25}{nm}$) as a function of gate voltage for fixed source drain voltage of \SI{0.9}{V}with linear regression fits to elucidate general trends.}
  \subfloat[Total Magnetization]{\label{fig:total-vs-VD} \includegraphics[width=0.32\textwidth]{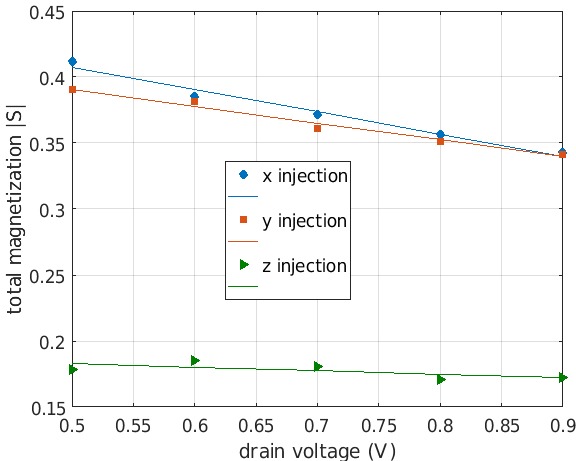}}\hfill
  \subfloat[$\theta$ angle]        {\label{fig:theta-vs-VD}\includegraphics[width=0.32\textwidth]{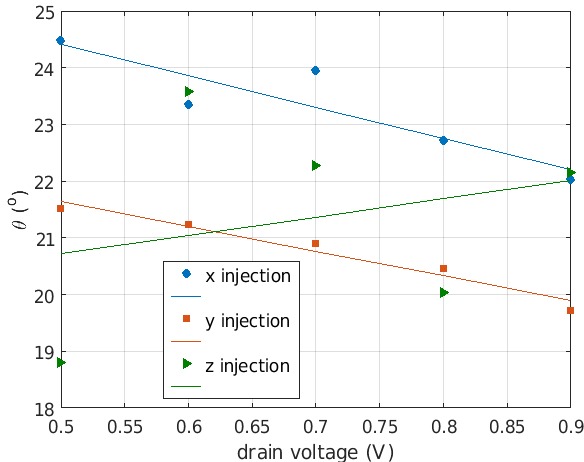}}\hfill
  \subfloat[$\phi$ angle]           {\label{fig:phi-vs-VD}    \includegraphics[width=0.32\textwidth]{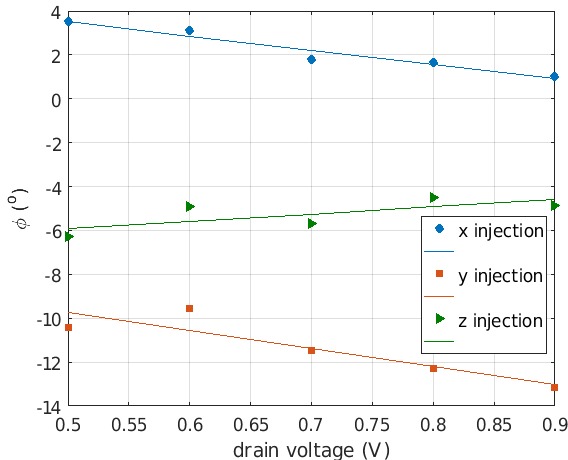}}\\
  \caption{\label{fig:VD-dependence}Magnetization at the left drain edge ($x=\SI{25}{nm}$) as a function of source-drain voltage for a fixed gate voltage of \SI{0.9}{V} with linear regression fits to elucidate general trends in the data.}
\end{figure*}

To assess the amount of coherent control of the magnetization possible without any external magnetic field, we explore the dependence of the magnetization on the respective voltages.

Fig.~\ref{fig:total-vs-VG} shows that the total magnetization at the drain edge decreases linearly with the gate voltage for $S_y$ and $S_z$-polarized injection but increases for $S_x$ injection. Fig.~\ref{fig:theta-vs-VG} further shows that the azimuthal angle $\theta$, indicating a rotation in the $S_x$-$S_y$ plane, increases linearly with the applied gate voltage for both $S_x$ and $S_y$-injection, while it decreases for $S_z$-injection. For $S_z$-injection the azimuthal rotation angle is greater but less sensitive to the gate voltage. Fig.~\ref{fig:phi-vs-VG} shows that the elevation angle $\phi$ is almost constant as a function of gate voltage for $S_x$ and $S_y$-injection, but decreases linearly for $S_z$-injection.  

Unlike for micron-size devices, where the rotation angle between the initial and final polarization can vary between $0$ and $180^\circ$ \cite{Bournel1997,Bournel1998,Bournel1998,Bournel2000,Bournel2001}, the dynamic range of the rotation angles for our nanometer-scale device is limited; in particular, $\theta$ is significantly less than $180^\circ$ indepedendent of the gate voltage.  However, the graphs suggest that the rotation angles are gate-voltage controllable and should give rise to modulations of the drain current, albeit smaller than what would be expected for micron-size devices.  On the positive side, the nanometer size of the device  significantly reduces the magnetization loss due to dephasing with the total magnetization at the drain ranging between 30\% and 40\% of the initial magnetization for $S_x$ or $S_y$ injection, and 17-22\% for $S_z$ injectiion at room temperature ($T=\SI{300}{K}$) in Fig.~\ref{fig:total-vs-VG}.

The total magnetization after a steady-state has been reached for different gate voltages, shown in Fig.~\ref{magVgate} in Appendix~\ref{Appx:Voltage}, suggests that lower gate voltages initially lead to faster magnetization decay, an effect that is most pronounced for $S_x$-polarized injection (see Fig.~\ref{magVgateA}).  A possible explanation for this effect is that larger gate voltages induce a high fringing electric field resulting in the electrons experiencing more acceleration. Electrons thus reach the gate faster and undergo fewer scattering events. However, the situation is more complicated for the final magnetization at the drain edge due to the magnetization recovery between the gate and drain. This recovery is more pronounced for lower gate voltages, possibly due to slower moving electrons experiencing less deceleration as they move toward the drain, giving the magnetization more time to recover.

\begin{figure}
  \includegraphics[width=.8\columnwidth]{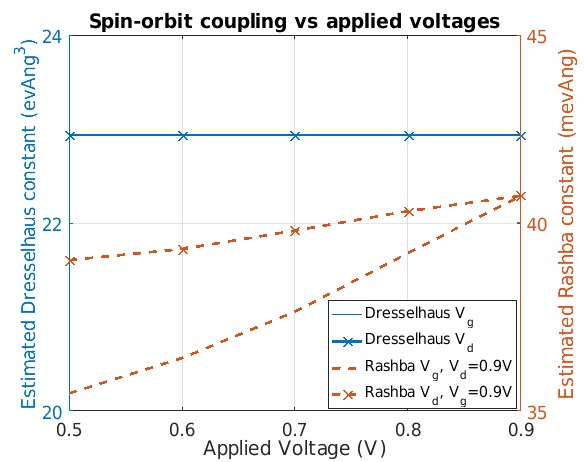}
  \caption{Dependence of Rashba and Dresselhaus coupling on source-drain and gate voltage.}
  \label{coupling-vs-voltage}
\end{figure}

The differences in the observed coherent rotation of the magnetization as a function of the applied voltages can be partially explained by changes in the strength of the spin-orbit coupling.  Fig.~\ref{coupling-vs-voltage} shows that the Rashba coupling constant increases with the applied voltages, while the Dresselhaus constant is independent of the voltage.  As the Rasbha and Dresselhaus Hamiltonians for a single electron spin correspond to rotations of the spin about orthogonal axes, changing their relative strengths will change the rotations experienced by the spins, although the ensemble picture is complicated and it is therefore not easy to relate the rotation of the magnetization vector representing the ensemble directly to changes in parameters such as the interaction strengths.  However, from a practical point of view it is sufficient if the simulations can correctly predict observable changes such as a gate-voltage dependent rotation of the net magnetization.

\subsection{Source-Drain Voltage Dependence}

The effect of varying the source-drain voltage on the total magnetization and its relative orientation at the left drain edge (after a steady-state has been reached) is shown in Fig.~\ref{fig:total-vs-VD} for a fixed gate voltage of \SI{0.9}{V}.  The latter value was chosen as the conventional \SI{25}{nm} gate length MOSFET for digital applications would have to operate at $V_{d}=V_{g}-V_\mathrm{th}=\SI{0.9}{V}$.  We observe a moderate decrease in the total magnetization at the drain edge with increasing source-drain voltage, especially for $S_x$ and $S_y$ injection.  For $S_z$ injection the total magnetization at the drain edge is significantly lower and appears to be effectively independent of the source-drain voltage.

Figs~\ref{fig:theta-vs-VD} and \ref{fig:phi-vs-VD} further suggest a linear decrease in both the azimuthal rotation angle $\theta$ and the elevation angle $\phi$ for $S_x$ and $S_y$ injection with increasing source-drain voltage if the gate voltage is fixed at $V_\mathrm{G}=\SI{0.9}{V}$, while for the $S_z$-injection both $\theta$ and $\phi$ appear to be effectively independent of the source-drain voltage.

The plots of the magnetization along the channel for different source-drain voltages and fixed gate voltage of $V_\mathrm{G}=\SI{0.9}{V}$, shown in Fig.~\ref{magVdrain} in Appendix~\ref{Appx:Voltage}, further suggest that the net magnetization along the channel is most affected by the drain voltage in the region between the gate and drain but is almost independent of the source-drain voltage in the region between the source and gate.  This makes sense physically as the electric field experienced by the electrons will be far more significantly influenced by the source-drain voltage in the region near the drain.  For $S_x$ and $S_y$ injection we observe a larger dip in the net magnetization in the gate region, partially offset by an increased recovery towards the drain edge, while for $S_z$ injection the net magnetization along the channel including in the region between the gate and drain appears to be almost independent of the source-drain voltage.

Although precise explanations for some of the observed effects and dependencies remain to be determined, the simulation results suggest that a certain amount of coherent control of spin transport in a realistic MOSFET-like device is possible by adjusting the gate and source-drain voltages alone, without any need for external magnetic fields even for \emph{nanoscale} devices operating at room temperature.  The sensitivity to voltage changes also exhibits a dependence on the injection polarization, which could be exploited in applications where it may be possible to dynamically change the polarization of the injected electrons.

\section{\label{sec:strain}Effect of Strain on Spin Transport}

Mechanical strain alters the amount of Dresselhaus coupling by changing the symmetry of the bulk crystal and the quantum well. Thus, application of strain could theoretically be a way to control the amount of spin-orbit coupling and the magnetization at the drain.  Alternatively, measuring the magnetization of the drain current could enable us to indirectly measure mechanical strain in the device and form the basis for a nanoscale mechanical strain sensor. To gain a better understanding of the effects of mechanical strain on the values of $\alpha_{br}$ and $\gamma$, we calculate the electronic bandstructures of our device as a function of mechanical strain. To this end we use the $\vec{k}\cdot\vec{p}$ method as it provides a good trade-off between efficiency and accuracy for bandstructure calculations for bulk semiconductors and heterostructures~\cite{M.Cardona1988,Neffati2012,Kane1957}.  

\subsection{Spin-Orbit Coupling of Strained Device}

\begin{figure*}
  \subfloat[Strain in direction {$[001]$}. ]{\includegraphics[width=0.32\textwidth]{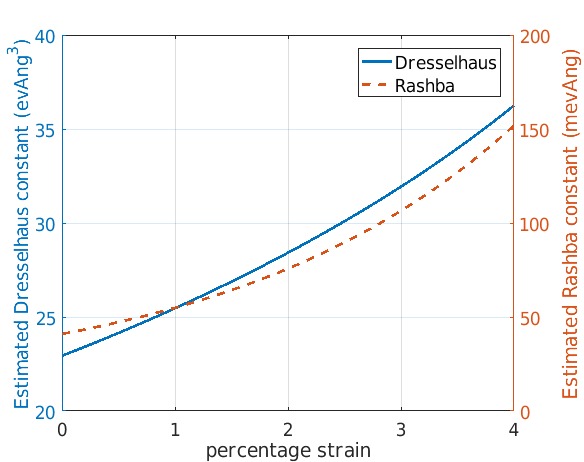}}\hfill
  \subfloat[Strain in direction {$[110]$}.]{\includegraphics[width=0.32\textwidth]{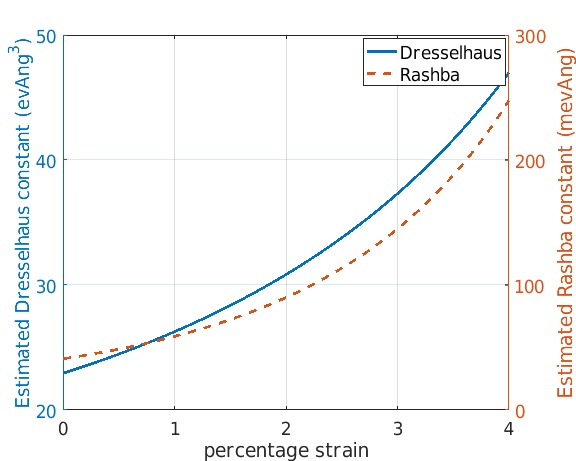}}\hfill
  \subfloat[Strain in direction {$[111]$}. ]{\includegraphics[width=0.32\textwidth]{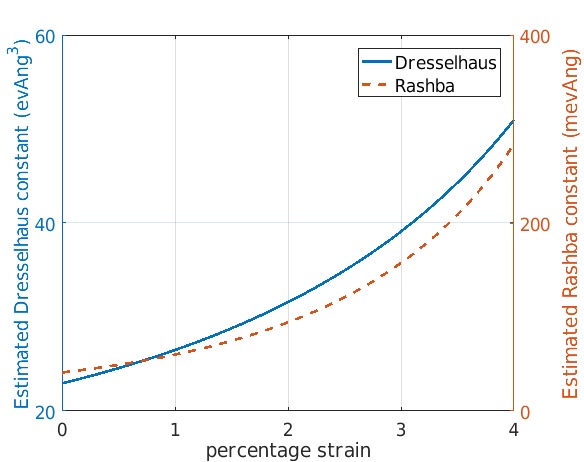}}
  \caption{\label{SpinVstrain}Spin-orbit coupling constants $\alpha_{br}$ and $\gamma$ as a function of strain ranging from $0\%$ to $4\%$ for three different strain directions. $\alpha_{br}$ has been calculated for $V_\mathrm{G} = \SI{0.9}{V}$ corresponding to an average electric field of $\SI{3.6e8}{V/m}$.}
  \subfloat[Total magnetization at drain]{\label{fig:total-vs-strain}\includegraphics[width=0.32\textwidth]{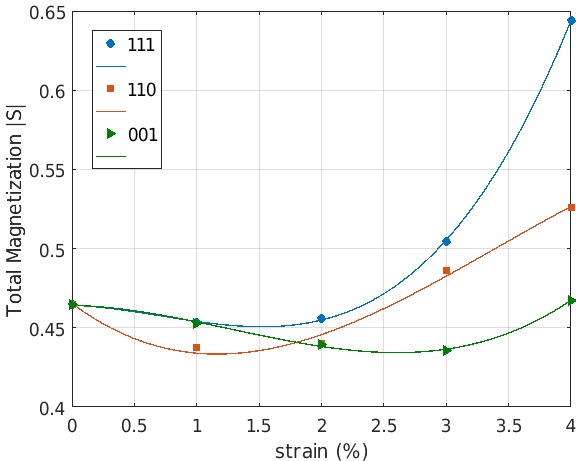}}\hfill
  \subfloat[Azimuthal angle at drain]{\label{fig:theta-vs-strain}\includegraphics[width=0.32\textwidth]{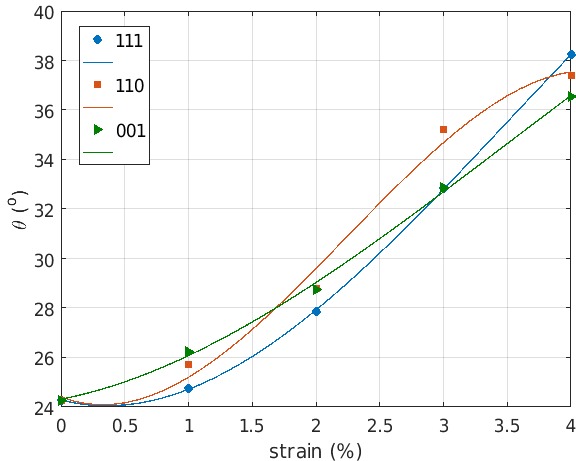}}\hfill
  \subfloat[Elevation angle at drain]{\label{fig:phi-vs-strain}\includegraphics[width=0.32\textwidth]{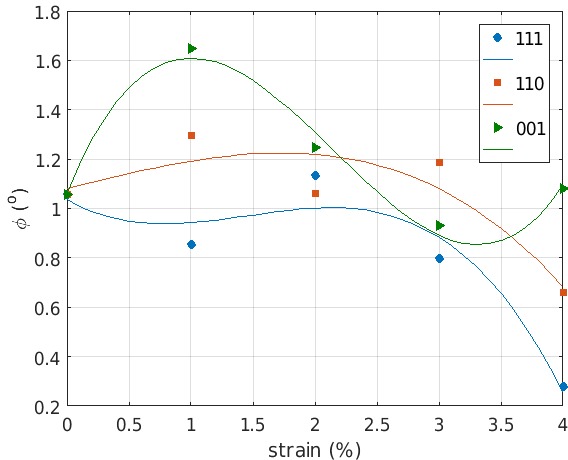}}
  \caption{\label{MagVstrain2}Steady state magnetization showing total magnetisation (a), at drain as a function of strain along different axis ($x$-injection, $V_D=V_G=\SI{0.9}{V}$) with nonlinear spline fits shown by lines to serve as a guide to the eye in elucidating trends in the data.}
\end{figure*}

The details of our $\vec{k}\cdot\vec{p}$ calculations are described in Appendix~\ref{Appx:kdotp}.  We integrate the method with our simulation techniques to include the effects of mechanical strain by first conducting $\vec{k}\cdot\vec{p}$ calculations for compressive strain in the $[001]$, $[110]$ and $[111]$ crystallographic directions. This is achieved by taking a small change in the lattice spacing $a_s$ of between $0$ and $4$\% of the unstrained lattice constant $a_0$. We then calculate the strain in the direction of the applied force as $e_{\|}=\tfrac{a_s}{a_0}-1$ and in the perpendicular direction as $e_{\bot}=-D_{[hkl]}e_{\|}$, where $h$, $k$ and $l$ are the Miller indices for the direction of the strain, and the coefficients $D_{[hkl]}$ are given by
\begin{subequations}
\begin{align}
  D_{[001]}&=\frac{2C_{12}}{C_{11}},\\
  D_{[110]}&=\frac{C_{11}+3C_{12}-2C_{44}}{C_{11}+C_{12}+2C_{44}},\\
  D_{[111]}&=\frac{2C_{11}+4C_{12}-4C_{44}}{C_{11}+2C{12}+4C{44}},
\end{align}
\end{subequations}
$C_{11}$,$C_{22}$ and $C_{44}$ being material dependant elastic constants.
The direction of the strain is accounted for by modifying the strain tensor $e_{hkl}$ as follows
\begin{equation}\label{strain_tensor(001)}
  e_{001}= \begin{pmatrix} e_{1} & 0 & 0 \\ 0 &e_{1} & 0 \\ 0 & 0 & e_{1} \end{pmatrix},
\end{equation}
\begin{equation}\label{strain_tensor(111)}
  e_{111}= \begin{pmatrix} e_2 & e_3 & e_3 \\ e_3 & e_2 & e_3 \\ e_3 & e_3 & e_2 \end{pmatrix},
\end{equation}
\begin{equation}\label{strain_tensor(110)}
  e_{110}= \begin{pmatrix} e_{4} & e_5 & 0 \\ e_5 & e_{4} & 0 \\ 0 & 0 & e_{1} \end{pmatrix},
\end{equation}
where
\begin{subequations}
\begin{align}
  e_1&=e_{\|},\\
  e_2&=\tfrac{1}{3}(e_{\bot}+2e_{\|}),\\
  e_3&=\tfrac{1}{3}(e_{\bot}-e_{\|}),\\
  e_4&=\tfrac{1}{2}(e_{\bot}+e_{\|}),\\
  e_5&=\tfrac{1}{2}(e_{\bot}+e_{\|}).
\end{align}
\end{subequations}

From this, we extract the new inter-band energies for the strained system $E_0'$, $E_1'$, $\Delta_0'$ and $\Delta_1'$ and use these to calculate the change in spin-orbit coupling parameters $\gamma$ and $\alpha_{br}$ via Eqs.~\eqref{Dresselhaus_const} and~\eqref{Rashba_const} derived in Appendix~\ref{Appx:kdotp}.  Finally, these new constants are inserted into our Monte Carlo simulation to investigate the effects of strain on the spin transport across the device.

\subsection{Effect of Strain on Magnetization}

Fig.~\ref{SpinVstrain} shows that the Dresselhaus and Rashba spin-orbit coupling constants both increase non-linearly with strain due to the change in the energy gap between the $\Gamma_{6c}$ and $\Gamma_{7v}$ bands at the Gamma point $E_0$.  To elucidate the effect of strain on spin transport and the steady state magnetization along the channel, the Monte Carlo device simulations were repeated for the same MOSFET device considered above using the strain-dependent values of the Rashba and Dresselhaus constants for different directions and strength of the strain.

Fig.~\ref{fig:total-vs-strain} shows a non-linear increase of total magnetization at the drain as a function of strain in the $[001]$ and $[110]$ directions, but no significant change for strain in the $[111]$ direction, while the small fluctuations of the elevation angle $\phi$ around $1^\circ$ for all three strain directions in Fig.~\ref{fig:phi-vs-strain} suggests that the elevation angle $\phi$ is not significantly affected by strain.  Fig.~\ref{fig:theta-vs-strain}, however, shows a non-linear increase of the azimuthal rotation angle $\theta$ with strain in all strain directions, including the $[111]$ direction, which shows no significant change in the total drain magnetization for increasing strain. This suggests that the azimuthal rotation angle of the magnetization vector may be the best measure of overall strain.

Fig.~\ref{MagVstrain} in Appendix~\ref{Appx:Voltage}  shows that both direction and magnitude of the strain have a direct impact on the amount of decay and recovery seen at the drain.  For strain in the $[001]$ direction with $S_x$ injection (Fig.~\ref{f:001}), a steady decrease in magnetization occurs when moving from the source to the gate, although at lower rates for stronger strain.  As before we observe some recovery of magnetization as we approach the drain.  Overall, the magnetization at the drain increases from a value of $0.36$ in the unstrained case to $0.64$ for $4.0\%$ strain.  The rate of spin recovery between the gate and drain, on the other hand, appears to decrease with increasing strain from a maximum of $0.13$ in the unstrained case to a minimum recovery of only $0.0$ for $4.0\%$ strain. Therefore, the increased magnetization at the drain is likely due to the decrease in the overall decay.

Fig.~\ref{f:110} shows similar results for strain in the $[110]$ direction. An increase in strain leads to a decrease in the amount of magnetization decay across the channel and an increase at the drain. As in the previous case, the rate of recovery decreases with increasing strain, with the unstrained case yielding a recovery of $0.13$ whilst the $4.0\%$ case only yields a recovery of $0.07$.  The result is notably different for increasing strain in the $[111]$ direction, shown in Fig.~\ref{f:111}. The increased spin-orbit coupling in this case has little effect on the total magnetization at the drain edge as the differences between the curves fall within the variance of the simulations. The magnetization still recovers at the drain, roughly by the same amount of $0.13$.

\section{Conclusions}

Ensemble Monte Carlo device simulations of electron-spin transport through a realistic structure of a \SI{25}{nm} gate length \chem{In_{0.3}Ga_{0.7}As} MOSFET show that the total magnetization and orientation of the magnetization (represented by the length and direction of the Bloch vector associated with the spin degrees of freedom) can be controlled via the gate voltage.  In this nanoscale transistor, the dynamic range of the rotation angles for the magnetization is significantly reduced compared to micron-size devices studied in previous works~\cite{Bournel1997, Bournel1998, Bournel2000, Bournel2001, MinShen2004}. Consequently, the magnetization loss due to dephasing becomes smaller and magnetization dependent modulations in the drain current can be still observable at room temperature. We observe a spin recovery between the gate and the drain due to a spin refocusing effect induced by the high electric fringing field at the gate~\cite{AynulIslam2011}. If experimentally verified, this recovery could be exploited in devices to increase spin polarization at the drain.

Our investigation into the effects of mechanical strain on the evolution of the magnetization shows that the spin transport is sensitive to strain. Larger strain leads to reduced magnetization decay when moving across the channel, dependent on the strain direction. The magnetization vector also undergoes a coherent rotation between the source and the drain due to spin-orbit coupling. Significantly, the azimuthal rotation angle exhibits a nonlinear increase with the amount of strain for all strain directions.

Although our focus in this work was understanding spin transport though a realistic \emph{nanoscale} MOSFET-like architecture at room temperature, and we did not wish to limit the scope of the simulation results by considering specific injection and detection mechanisms in the source/drain, these issues play important role in practice.  In particular, the possibility of room temperature operation will depend on whether we can achieve sufficiently high injection and detection efficiencies to measure the predicted changes, which may be challenging if these changes are small.  However, as we are working in the steady-state transport regime, the signal-to-noise ratio of measurements can potentially be increased by measuring the output polarization over an extended period of time.  Furthermore, while it may be difficult to detect small changes in the net magnetization at the drain, especially if the injection and detection efficiencies are not precisely known, the angles by which the polarization is rotated and, in particular, the voltage and strain dependence of these angles are independent of the injection and detection efficiency.  Optimization of the device geometry such as channel length, oxide thickness, and doping profile to maximize the magnetization recovery may further boost the attainable signal-to-noise ratio.

\begin{acknowledgments}
This research was funded by the S\^er Cymru National Research Network for Advanced Engineering and Materials (grant NRN~$082$). SS also acknowledges funding from a Royal Society Leverhulme Trust Senior Research Fellowship. We would like to thank J. Fabian (Regensburg University) for helpful comments and suggestions.
\end{acknowledgments}

\bibliography{paper}

\appendix

\section{The $\mathbf{k\cdot{}p}$ Method}\label{Appx:kdotp}

\begin{figure*}
\centering
\null\hfill
  \subfloat[Unstrained case.]{\includegraphics[width=0.49\textwidth]{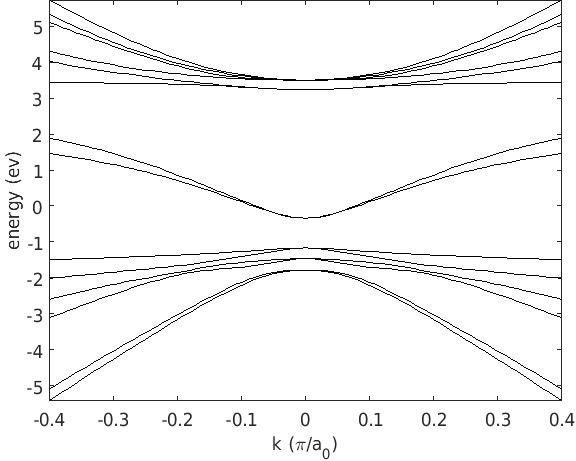}}\hfill
  \subfloat[4\% strain in direction {$[001]$}.]{\includegraphics[width=0.49\textwidth]{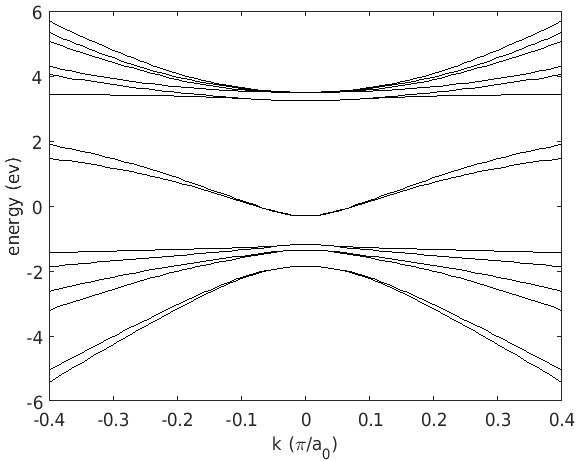}}\hfill\null\\
\null\hfill
  \subfloat[4\% strain in direction {$[110]$}.]{\includegraphics[width=0.49\textwidth]{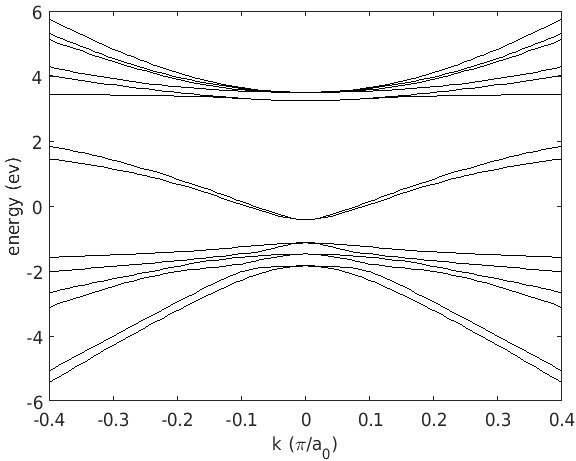}}\hfill
  \subfloat[4\% strain in direction {$[111]$}.]{\includegraphics[width=0.49\textwidth]{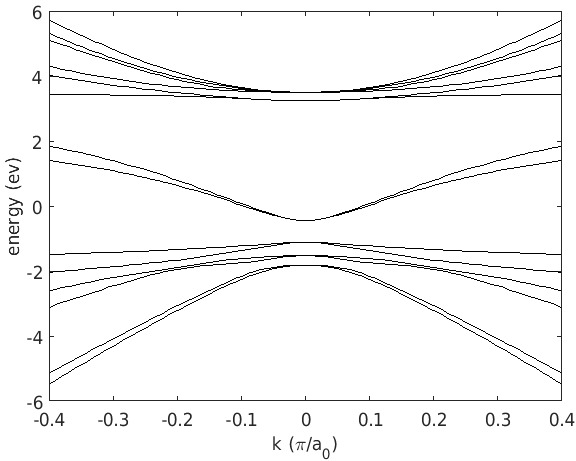}}\hfill\null\\
  \caption{\label{fig:bands}Band structures for \chem{In_{0.3}Ga_{0.7}As} calculated using $\vec{k}\cdot\vec{p}$ method with different mechanical strain.}
\end{figure*}

The method consists of calculating the band structure in the vicinity of a given point in reciprocal-space for which the band structure is known using the perturbation theory. First proposed by Kane in 1957~\cite{Kane1957}, including only the lowest conduction band ($\Gamma _{6c}$), the light hole, heavy hole ($\Gamma_{8v}$) and split off ($\Gamma_{7v}$) valence bands, it has since been extended to include up to $40$ bands and can be easily adapted to include the effects of spin-orbit coupling and strain~\cite{Neffati2012}. In our case we only need to consider $7$ bands ($14$ with the inclusion of a spin), which is achieved by including the next two conduction bands (the doublet $\Gamma_{7c}$ and quadruplet $\Gamma_{8c}$ bands).

The Hamiltonian for this system is constructed following the procedure used by Pfeffer and Zawadzki~\cite{Pfeffer1990,Pfeffer1996}. Starting with the general Schr{\"o}dinger equation for an electron wave function $\psi_{k}$ with wavevector $\vec{k}$,
\begin{equation} \label{schro}
  \begin{split}
    [H+ V(r)]\psi_{k}(r) = E(k)\psi_{k}(r), & \\
    H =\frac{p^2}{2m_0}+\frac{\hbar}{4m_0^2c^2}(\sigma \times \nabla V) \cdot \vec{p},&
  \end{split}
\end{equation}
where $\sigma$ represents the vector of Pauli matrices and $V(r)$ is a periodic potential, we look for solutions to Eq.~\eqref{schro} of Luttinger-Kohn form~\cite{Luttinger1955},
\begin{equation}\label{lutt-kohn}
  \psi_{k}^m(r) = e^{ik\cdot r}\sum_{l}c_{l}^{m}u_l(r),
\end{equation}
where the summation is over all bands and the index $m$ indicates the band of interest. The L-K periodic amplitudes satisfy Eq.~\eqref{schro} at a band's extremity ($k=0$ in our case) and are orthonormal such that $\left(\frac{1}{\Omega}\right)\braket{u_{i'}}{u_l}=\delta_{l'l}$ where $\Omega$ is the unit cell over which the integration is taking place. Thus, by substituting Eq.~\eqref{lutt-kohn} into Eq.~\eqref{schro}, multiplying the left hand side by $\left(\tfrac{1}{\Omega}\right)u(r)^*$ and integrating over the unit cell, we obtain
\begin{equation}
  \sum_l \left[(E^{(l)}+\frac{\hbar^2k^2}{2m_0}-E)\delta_{l'l}+\frac{\hbar}{m_0}k\cdot p_{l'l}+H_{l'l}^{s.o}\right]c_l^m=0,
\end{equation}
where the index $l'$ runs over all the bands, $E^{(l)}$ is the band energy and $p_{l'l}$ are the inter-band matrix elements.

Finally, we must account for the effects of the interactions with far-level bands. This is achieved by using the method developed by Lowdin~\cite{Lowdin1951,Chuang1995}, which uses a combination of perturbation theory for the quasi-degenerate levels and the far-levels. Several additional Hamiltonian parameters are required to account for this. We use the procedure outlined in Pfeffer and Zawadzki~\cite{Pfeffer1996}, which includes the diagonal contributions to the conduction $\Gamma_{6_c}$ band, the contributions to the Luttinger valence $\gamma^l$ parameters, and the linear $k$ terms, while all other far-level terms are neglected for simplicity. We use $\gamma_i$ to represent the modified Luttinger parameters, where the $\vec{k}\cdot\vec{p}$ interaction of the $\Gamma_{8v}$ level with the $\Gamma_{6c}$, $\Gamma_{8c}$ and $\Gamma_{7c}$ levels are subtracted as they are included explicitly in the matrix.

\begin{table}[t]
  \begin{tabular}{c|c|c}
  Band & Basis function & Energy\\\hline
  $\Gamma_{6c}$ & $\ket{S\uparrow} $                                                                                                    & $0$\\
                & $\ket{S\downarrow}$                                                                                                             & $0$\\
  $\Gamma_{8v}$ & $-\tfrac{1}{2}\ket{(X+iY)\uparrow}$                                                                       & $-E_0$\\
                & $ \tfrac{1}{2}\ket{(X+iY)\downarrow}$                                                                               & $-E_0$\\
                & $\sqrt{\tfrac{2}{3}}\ket{Z\uparrow}   - \tfrac{1}{\sqrt{6}}\ket{(X+iY)\downarrow}$      & $-E_0$\\
                & $\sqrt{\tfrac{2}{3}}\ket{Z\downarrow} + \tfrac{1}{\sqrt{6}}\ket{(X+iY)\uparrow}$      & $-E_0$\\
  $\Gamma_{7v}$ & $-\tfrac{1}{\sqrt{3}}\ket{Z\uparrow}   - \tfrac{1}{\sqrt{3}}\ket{(X+iY)\downarrow}$  & $-(E_0+\Delta_0)$\\
                & $ \tfrac{1}{\sqrt{3}}\ket{Z\downarrow} - \tfrac{1}{\sqrt{3}}\ket{(X+iY)\uparrow}$       & $-(E_0+\Delta_0)$\\
  $\Gamma_{8c}$ & $-\tfrac{1}{2}\ket{(X'+iY')\uparrow}$                                                                     & $E_1-E_0+\Delta_1$\\
                & $ \tfrac{1}{2}\ket{(X'+iY')\downarrow}$                                                                             & $E_1-E_0+\Delta_1$\\
                & $\sqrt{\tfrac{2}{3}}\ket{Z'\uparrow}   - \tfrac{1}{\sqrt{6}}\ket{(X'+iY')\downarrow}$   & $E_1-E_0+\Delta_1$\\
                & $\sqrt{\tfrac{2}{3}}\ket{Z'\downarrow} + \tfrac{1}{\sqrt{6}}\ket{(X'+iY')\uparrow}$   & $E_1-E_0+\Delta_1$\\
  $\Gamma_{7c}$ & $-\tfrac{1}{\sqrt{3}}\ket{Z'\uparrow}   - \tfrac{1}{\sqrt{3}}\ket{(X'+iY')\downarrow}$ & $E_1-E_0$\\
                & $ \tfrac{1}{\sqrt{3}}\ket{Z'\downarrow} - \tfrac{1}{\sqrt{3}}\ket{(X'+iY')\uparrow}$   & $E_1-E_0$\\
  \end{tabular}
  \caption{\label{t:bandfunc}Basis functions/energies for $14$-band $\vec{k}\cdot\vec{p}$ model~\cite{JaroslavFabian2007}}
\end{table}

With this in mind, we construct the Hamiltonian for the system by careful selection of basis functions for each band of interest, defined in Table~\ref{t:bandfunc}. These basis functions are chosen deliberately to exploit orbital symmetries as for III-V semiconductors the $\Gamma_{6c}$ conduction band consists of electrons from $s$-type orbitals while the valence bands $\Gamma_{8v}$ and $\Gamma_{7v}$  and the two remaining conduction bands $\Gamma_{8c}$ and $\Gamma_{7c}$ consist of electrons from $p$-type orbitals and only inter-band interactions of opposite parity will produce non-zero results. Thus we greatly reduce the number of parameters needed to calculate the band structure such that the only non-zero momentum matrix elements are given by
\begin{subequations}
  \begin{align*}
    P_0     & = h' \bra{ S\sigma}p_x\ket{X\sigma}
              = h' \bra{S\sigma}p_x\ket{Y\sigma}\\
            & = h' \bra{S\sigma}p_x\ket{Z\sigma},\\
    P_1     & = ih' \bra{X'}p_x\ket{S}
              = ih' \bra{Y'}p_y\ket{S}
              = ih' \bra{Z'}k_z\ket{S},\\
      Q     & = h' \bra{X}p_y\ket{Z'}
              =-h'\bra{X}p_y\ket{Z'},\\
  \Delta^-  & = -\tfrac{3h'}{4m_*}\bra{X}[(\nabla V_0) \times p]_y\ket{Z'}\\
            & =  \tfrac{3h'}{4m_*}\bra{Z}[(\nabla V_0) \times p]_y\ket{X'}
  \end{align*}
\end{subequations}
with $h' = \hbar/m_0$ and $m_* = m_0 c^2$. These elements can be taken as phenomenological parameters that can be determined experimentally.

With these assumptions the Hamiltonian $H_{0}$ for this $7$-band system is given by the matrix~\cite{Pfeffer1996,Bahder1990}
\begin{equation}
  \label{systemham}
  H_{0} = \begin{pmatrix}
            H_c & H_{cv} &H_{cc'}\\ H_{vc} & H_{v} & H_{vc'}\\ H_{c'c} & H_{c'v} & H_{c'}
          \end{pmatrix}
\end{equation}
with elements 
\begin{subequations}
\begin{align}
  H_c    &= \begin{pmatrix}
              \frac{\hbar^2k^2}{2m_0}+V(r) & 0 \\
              0 & \frac{\hbar^2k^2}{2m_0}+V(r)
            \end{pmatrix},\\
  H_{vc} &= \frac{1}{\sqrt{6}}
            \begin{pmatrix}
              -\sqrt{3} P_0 k_- & 0\\
              -2iP_0\pdv{y}          & -P_0 k_-\\
              P_0 k_+           & -2iP_0\pdv{y} \\
              0                 & \sqrt{3}P_0 k_+\\
              i\sqrt{2}P_0\pdv{y}  & -\sqrt{2}P_0 k_-\\
             7 -\sqrt{2}P_0 k_+  & -i\sqrt{2}P_0\pdv{y}
            \end{pmatrix},
\end{align}
\begin{widetext}
\begin{align}
  H_{cv}&=
    \begin{pmatrix}
      -\frac{1}{\sqrt{2}}P_0 k_+&-i\sqrt{\frac{2}{3}}P_0\pdv{y}&\frac{1}{\sqrt{6}}P_0 k_-&0&-\frac{i}{\sqrt{3}}P_0\pdv{y}&-\frac{1}{\sqrt{3}}P_0 k_- \\
      0 & -\frac{1}{\sqrt{6}}P_0 k_+&-i\sqrt{\frac{2}{3}}P_0\pdv{y}&\frac{1}{\sqrt{2}}P_0 k_-&-\frac{1}{\sqrt{3}}P_0 k_+&\frac{i}{\sqrt{3}}P_0\pdv{y}
    \end{pmatrix},\\
  H_{cc'}&=
    \begin{pmatrix}
      -\frac{i}{\sqrt{2}}P_1k_+&\sqrt{\frac{2}{3}}P_1\pdv{y}&\frac{i}{\sqrt{6}}P_1k_- &0&\frac{1}{\sqrt{3}}P_1\pdv{y}&-\frac{i}{\sqrt{3}}P_1k_-\\
      0&-\frac{i}{\sqrt{6}}P_1k_+&\sqrt{\frac{2}{3}}P_1\pdv{y}&\frac{i}{\sqrt{2}}P_1k_- &-\frac{i}{\sqrt{3}}P_1k_+&-\frac{i}{\sqrt{3}}P_1\pdv{y}\\
    \end{pmatrix},\\
      H_v &=
    \begin{pmatrix}
      (D+G)+E_v&0&-S^-&R&-\frac{S^-}{\sqrt{2}}&\sqrt{2}R \\
      0&(D+G)+E_v&R^+&S^+&-\sqrt{2}R^+&-\frac{S^+}{\sqrt{2}}\\
      -S^+&R&(D-G)+E_v&S^-&-\sqrt{2}G&\sqrt{\frac{3}{2}}S^-\\
      R^+&S^-&S^+&(D-G)+E_v&\sqrt{\frac{3}{2}}S^+&\sqrt{2}G \\
      \frac{-S^+}{\sqrt{2}}&-\sqrt{2}R&-\sqrt{2}G^+&\sqrt{\frac{3}{2}}S&D+E_v-\Delta_0&0 \\
      \sqrt{2}R&\frac{-S^-}{\sqrt{2}}&\sqrt{\frac{3}{2}}S^+&\sqrt{2}G^+&0&D+E_v-\Delta_0
    \end{pmatrix}\\
  H_{vc'}&=
    \begin{pmatrix}
      \frac{i}{3}\Delta^-&\frac{i}{\sqrt{3}}Qk_+&\frac{1}{\sqrt{3}}Q\pdv{y}&0&-\frac{i}{\sqrt{6}}Qk_+&-\sqrt{\frac{2}{3}}Q\pdv{y}\\
      -\frac{i}{\sqrt{3}}Qk_-&\frac{i}{3}\Delta^-&0&\frac{1}{\sqrt{3}}Q\pdv{y}&0&\frac{i}{\sqrt{2}}Qk_+\\
      -\frac{1}{\sqrt{3}}Q\pdv{y}&0&\frac{i}{3}\Delta^- & -\frac{i}{\sqrt{3}}Qk_+&-\frac{i}{\sqrt{2}}Qk_-&0\\
      0&-\frac{1}{\sqrt{3}}Q\pdv{y}&\frac{i}{\sqrt{3}}Qk_-&\frac{i}{3}\Delta_-&-\sqrt{\frac{2}{3}}Q\pdv{y}&\frac{i}{\sqrt{6}}Qk_-\\
      \frac{i}{\sqrt{6}}Qk_-&0&\frac{i}{\sqrt{2}}Qk_+&\sqrt{\frac{2}{3}}Q\pdv{y}&-\frac{2i}{3}\Delta^-&0\\
      \sqrt{\frac{2}{3}}Q\pdv{y}&-\frac{i}{\sqrt{2}}Qk_-&0&-\frac{i}{\sqrt{6}}Qk_+&0&-\frac{2i}{3}\Delta^-
    \end{pmatrix},
\end{align}
\begin{align}
  H_{c}'&=
    \begin{pmatrix}
      \frac{\hbar^2\textbf{k}^2}{2m_0}+E_{c'}+\Delta_1&0&0&0&0&0\\
      0&\frac{\hbar^2\textbf{k}^2}{2m_0}+E_{c'}+\Delta_1&0&0&0&0\\
      0&0&\frac{\hbar^2\textbf{k}^2}{2m_0}+E_{c'}+\Delta_1&0&0&0\\
      0&0&0&\frac{\hbar^2\textbf{k}^2}{2m_0}+E_{c'}+\Delta_1&0&0\\
      0&0&0&0&\frac{\hbar^2\textbf{k}^2}{2m_0}+E_{c'}&0\\
      0&0&0&0&0&\frac{\hbar^2\textbf{k}^2}{2m_0}+E_{c'}
    \end{pmatrix}.
\end{align}
\end{widetext}
\end{subequations}

This Hamiltonian was modified from the bulk case using the approach in \cite{Winkler2003,Bastard1988,RafaelS.Calsaverini2008} to account for confinement in the $y$-direction due to the quantum well by replacing $K_z\to-i\pdv{y}$ and taking all Luttinger parameters and momentum matrix elements to be $y$-dependent. 

Taking $E_v=V(r)-E_0$, $E_{c'}=V(r)+E_1$, $k_\pm = k_x\pm ik_y$, $k^2_{\parallel}=k_{x}^2+k_{y}^2$ and $k^2=k^2_\parallel+\pdv[2]{y}$, $H_{cc'}$ and $H_{c'v}$ are obtained by the transposition of $H_{c'c}$ and $H_{vc'}$, respectively, with the following substitutions: $k_\pm \to k_\mp$, $P_1\to-P_1$, $Q\to-Q$ and $\Delta^- \to -\Delta^-$, where
\begin{subequations}
\begin{align}
  D&=\tfrac{\hbar^2}{2m_0}(\gamma_1k^2_\parallel-\pdv{y}\gamma_1\pdv{y}),\\
  G&=\tfrac{\hbar^2\gamma_2}{2m_0}(\gamma_2k^2_\parallel+2\pdv{y}\gamma_2\pdv{y}),\\
  S^\pm&= \tfrac{\hbar^2\gamma_3}{m_0}(-i\sqrt{3}k_{\pm}\pdv{y})\,\\
  R&= \tfrac{\hbar^2}{2m_0}\left[-\sqrt{3}\gamma_2(k_x^2-k_y^2)+i2\sqrt{3}\gamma_3k_xk_y\right].
\end{align}
\end{subequations}

The $\gamma_i$ represent modified Luttinger parameters from which the $\vec{k}\cdot\vec{p}$  interaction of the $\Gamma_{8v}$ level with the $\Gamma_{6c}$, $\Gamma_{8c}$ and $\Gamma_{7c}$ levels has been subtracted as it is included explicitly in the matrix. The $\gamma_i$ are given by
\begin{subequations}\label{modlut}
\begin{align}
  \gamma_1&=\gamma_{1}^L+\tfrac{E_{p_0}}{3E_0}-\tfrac{E_Q}{3(E_0'-E_0)}-\tfrac{E_Q}{3(E_0'-E_0+\Delta_1)},\\
  \gamma_2&=\gamma_{2}^L+\tfrac{E_{p_0}}{6E-0}-\tfrac{E_Q}{6(E_0'-E_0)},\\
  \gamma_3&=\gamma_{3}^L+\tfrac{E_{p_0}}{6E-0}-\tfrac{E_Q}{6(E_0'-E_0)}.
\end{align}
\end{subequations}

To adapt this Hamiltonian to account for the effects of mechanical strain on the system, we follow Pikus and Bir~\cite{Chuang1995}. The method consists of adding an extra term to each element of the unstrained  Hamiltonian created by replacing $k_xk_y$ (and it's circular permutations) with the component of the strain tensor $\epsilon_{xy}$ and the Luttinger parameters with deformation potentials. Thus, the total system Hamiltonian is given by $H=H_{0} +H_{\epsilon}$.  The strain Hamiltonian $H_\epsilon$ for the seven band system described above is given by~\cite{Bahder1990,Chuang1995}
\begin{equation}\label{strainham}
  H_{\epsilon}=
    \begin{pmatrix}
      H_c      & H_{cv} & 0_{2\times6}\\
      H_{vc} & H_{v}  & 0_{6\times6}\\
      0_{6\times2} & 0_{6\times6} & 0_{6\times6}
    \end{pmatrix}
\end{equation}
where $H_{vc}=H_{cv}^\dag$. Setting $A_{\epsilon} = a_c(\epsilon_{xx}+\epsilon_{yy}+\epsilon_{zz})$, $v = \tfrac{P_0}{\sqrt{6}}\sum_j(\epsilon_{xj}-i\epsilon_{yj})$ and $u =\tfrac{P_0}{\sqrt{3}}\sum_j\epsilon_{zj}$  we have
\begin{align*}
H_c     &= \begin{pmatrix} A_{\epsilon}&0 \\ 0&A_{\epsilon} \end{pmatrix},\\
H_{cv}  &= \begin{pmatrix} \sqrt{3}v^\dagger&-\sqrt{2}u&-v&0&-u&-\sqrt{2}v\\
                           0&v^\dagger&-\sqrt{2}u&-\sqrt{3}v&\sqrt{2}v^\dagger&u \end{pmatrix}
\end{align*}
and setting
\begin{align*}
  D_{\epsilon} &=-a_v(\epsilon_{xx}+\epsilon_{yy}+\epsilon_{zz}),\\
  G_{\epsilon} &=-b(\epsilon_{xx}+\epsilon_{yy}-2\epsilon_{zz})/2,\\
  S_{\epsilon} &=-d(\epsilon_{xz}-i\epsilon_{yz}),\\
  R_{\epsilon} &=\tfrac{\sqrt{3}}{2}b(\epsilon_{xx}-\epsilon_{yy})-id\epsilon_{xy}
\end{align*}
we have
\begin{widetext}
\begin{align*}
  H_v &=
    \begin{pmatrix}
      (D_{\epsilon}+G_{\epsilon})&0&-S_{\epsilon}&R_{\epsilon}&-\frac{S_{\epsilon}}{\sqrt{2}}&\sqrt{2}R_{\epsilon} \\
      0&(D_{\epsilon}+G_{\epsilon})&R_{\epsilon}^\dagger&S_{\epsilon}^\dagger&-\sqrt{2}R_{\epsilon}^\dagger&-\frac{S_{\epsilon}^\dagger}{\sqrt{2}}\\
      -S_{\epsilon}^\dagger&R_{\epsilon}&(D_{\epsilon}-G_{\epsilon})&S_{\epsilon}&-\sqrt{2}G_{\epsilon}&\sqrt{\frac{3}{2}}S_{\epsilon}\\
      R_{\epsilon}^\dagger&S_{\epsilon}&S_{\epsilon}^\dagger&(D_{\epsilon}-G_{\epsilon})&\sqrt{\frac{3}{2}}S_{\epsilon}^\dagger&\sqrt{2}G_{\epsilon} \\
      \frac{-S_{\epsilon}^\dagger}{\sqrt{2}}&-\sqrt{2}R_{\epsilon}&-\sqrt{2}G_{\epsilon}^\dagger&\sqrt{\frac{3}{2}}S_{\epsilon}&D_{\epsilon}-\Delta_0&0 \\
      \sqrt{2}R_{\epsilon}&\frac{-S_{\epsilon}}{\sqrt{2}}&\sqrt{\frac{3}{2}}S_{\epsilon}^\dagger&\sqrt{2}G_{\epsilon}^\dagger&0&D_{\epsilon}-\Delta_0
    \end{pmatrix}.
\end{align*}
\end{widetext}

As the influence of strain on the $p$-type conduction bands ($\Gamma_{7c}$ and $\Gamma_{8c}$) is currently unknown, the $p$-type conduction band hydrostatic deformation potential ($a_{\Gamma_{7c}}$ and $a_{\Gamma_{8c}}$) and $p$-type CB shear deformation potential ($b_{\Gamma_{7c}}$ and $b_{\Gamma_{8c}}$) are neglected.  The results of bandstructure calculations for different levels and types of strain are shown in Fig.~\ref{fig:bands}.

With the parameters gained from these calculations, we estimate the Dresselhaus and Rashba constants \cite{Winkler2003,JaroslavFabian2007,J.M.Jancu2005}
\begin{equation}\label{Dresselhaus_const}
  \begin{aligned}
    \gamma=\dfrac{4P_0P_1Q}{3}\qty[\dfrac{1}{E_0E_1}-\dfrac{1}{(E_0+\Delta_0)(E_1+\Delta_1)}]\\
    -\dfrac{4P_0^2Q\Delta^-}{9E_0(E_0+\Delta_0)}\qty[\dfrac{2}{E_1}+\dfrac{1}{E_1+\Delta_1}]\\
    -\dfrac{4P_1^2Q\Delta^-}{9E_1(E_1+\Delta_1)}\qty[\dfrac{1}{E_0}+\dfrac{2}{E_1+\Delta_0}],
  \end{aligned}
\end{equation}
\begin{equation}\label{Rashba_const}
  \begin{aligned}
    \alpha_{br}=\dfrac{1}{3}\qty[\dfrac{P_0^2}{(E_0+\Delta_0)^2}-\dfrac{P_0^2}{E_0^2}+\dfrac{P_1^2}{E_1^2}-\dfrac{P_1^2}{(E_1+\Delta_1)^2}]\dv{V_{ext}}{y}\\
    -\dfrac{2P_1P_0\Delta^-}{9}\left[\dfrac{1}{E_1(E_1+\Delta_1)^2}-\dfrac{1}{E_0^2(E_1+\Delta_1)}\right.\qquad\\
    \left.-\dfrac{2}{E_1(E_0+\Delta_0)^2}+\dfrac{2}{E_1^2(E_0+\Delta_0)}\right]\dv{V_{ext}}{y}
  \end{aligned}
\end{equation}
and by extension the change in spin-orbit coupling with respect to mechanical strain. Values for an unstrained \chem{In_{0.3}Ga_{0.7}As} quantum well of $\alpha_{br}=40.94\si{\milli\electronvolt\angstrom}$ and $\gamma=22.94 \si{\electronvolt\angstrom^3}$ obtained from these equations compare well with both theoretical 40-band tight-binding calculations \cite{J.M.Jancu2005,JaroslavFabian2007} and experimental measurements \cite{W.Knap1996}. Our values also compare well to the self-consistent approaches conducted by \cite{RafaelS.Calsaverini2008,Dettwiler2014,JiyongFu2016}. 

\clearpage
\newpage
\onecolumngrid
\section{Voltage and Strain Dependence plots}\label{Appx:Voltage}
\begin{figure}[h]
\begin{center}
 \subfloat[Injection $x$-polarized.]{\label{magVgateA}\includegraphics[width=0.3\textwidth]{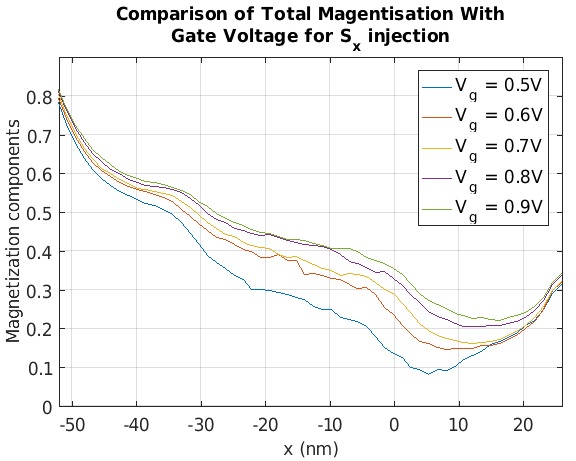}}\hfill
 \subfloat[Injection $y$-polarized.]{\label{magVgateB}\includegraphics[width=0.3\textwidth]{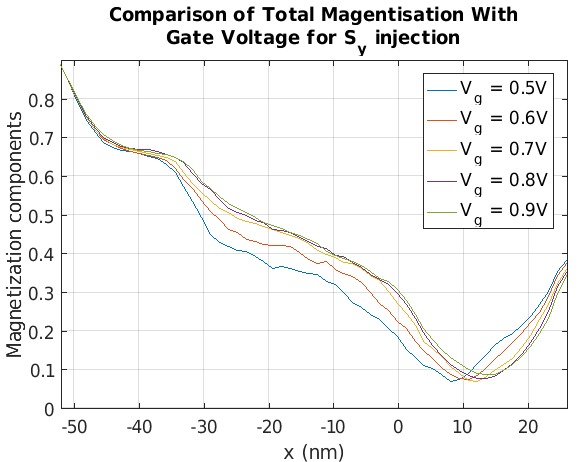}}\hfill
 \subfloat[Injection $z$-polarized.]{\label{magVgateC}\includegraphics[width=0.3\textwidth]{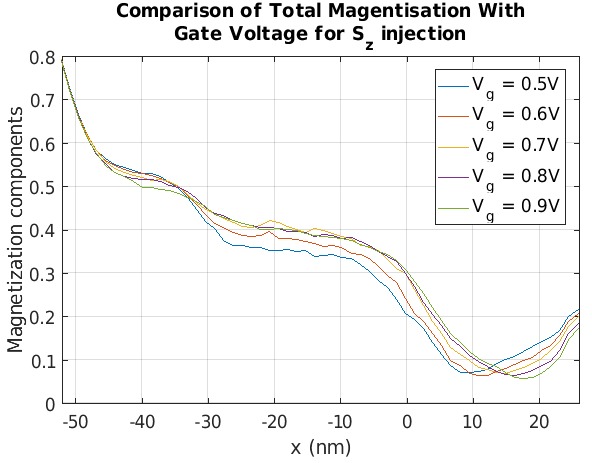}}\\
 \caption{\label{magVgate}Total magnetization versus position along the channel after steady-state has been reached at different gate voltages (V$_\mathrm{g}$) and a fixed source-drain voltage (V$_\mathrm{d}$) of \SI{0.9}{V}.}

 \subfloat[Injection $S_x$-polarized.]{\includegraphics[width=0.32\textwidth]{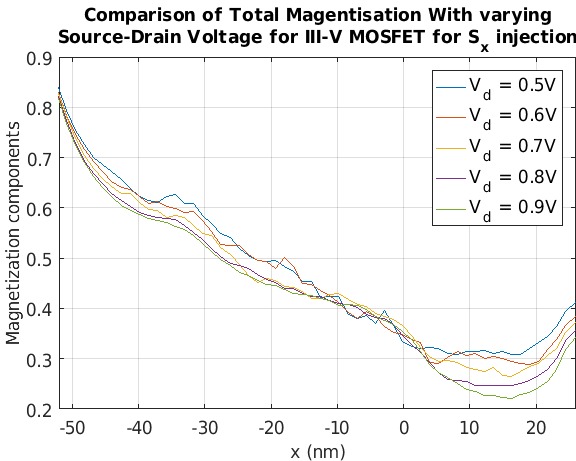}}\hfill
 \subfloat[Injection $S_y$-polarized.]{\includegraphics[width=0.32\textwidth]{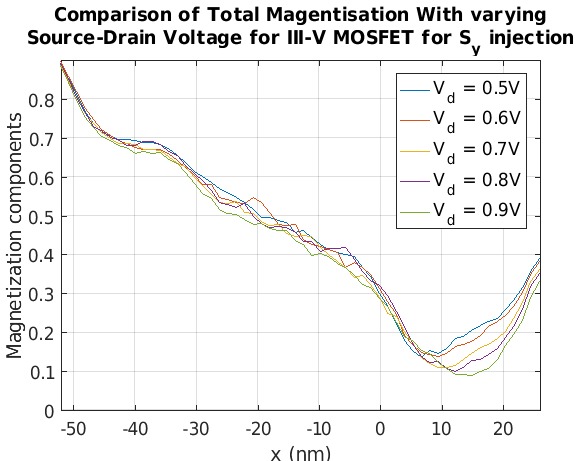}}\hfill
 \subfloat[Injection $S_z$-polarized.]{\includegraphics[width=0.32\textwidth]{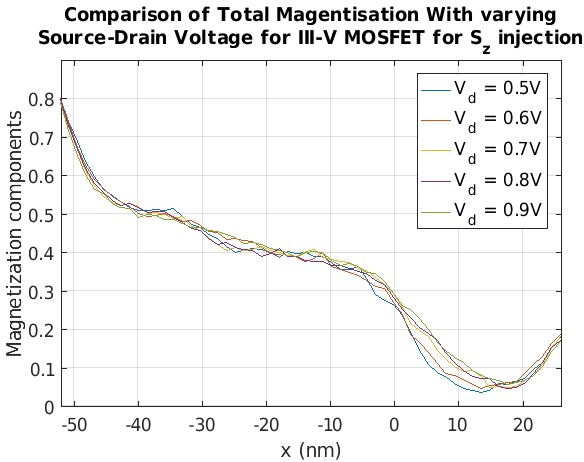}}\\
 \caption{\label{magVdrain}Total magnetization versus position along the channel after steady state has been reached at different source-drain voltages and a fixed gate voltage of \SI{0.9}{V}.}

 \subfloat[Strain in direction {$[001]$}.]{\label{f:001}\includegraphics[width=0.32\textwidth]{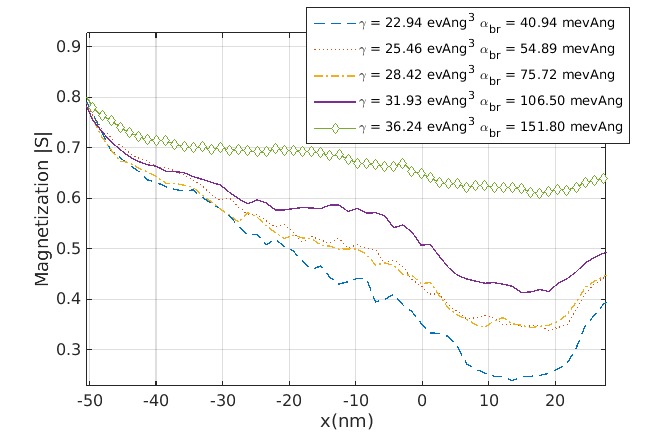}}\hfill
 \subfloat[Strain in direction {$[110]$}.]{\label{f:110}\includegraphics[width=0.32\textwidth]{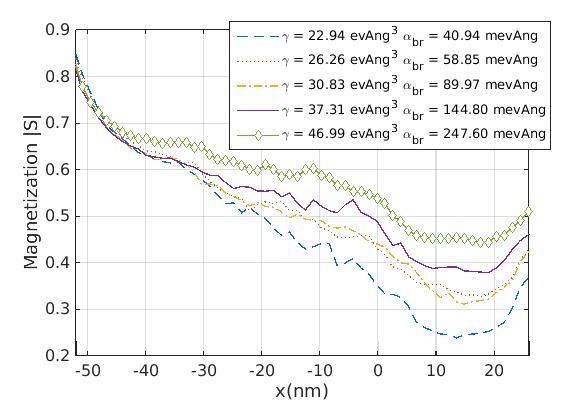}}\hfill
 \subfloat[Strain in direction {$[111]$}.]{\label{f:111}\includegraphics[width=0.32\textwidth]{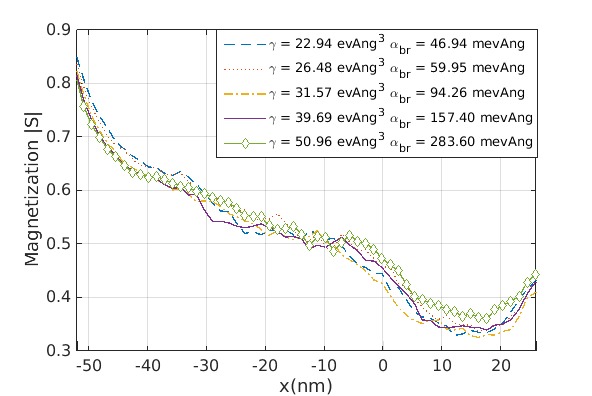}}\\
 \caption{\label{MagVstrain}Magnetization along the device channel vs. strain ranging from $0\%$ to $4\%$ for three different strain directions, taken after a steady state was reached at $t=\SI{8}{ps}$ for a gate and source-drain voltage of $\SI{0.9}{V}$ for $x$-injection for indicated Dresselhaus ($\gamma$) and Rashba ($\alpha_{br}$) spin-orbit coupling constants.}
\end{center}
\end{figure}
\end{document}